\documentclass[prc,aps,twocolumn,showpacs,amssymb,superscriptaddress]{revtex4}

\usepackage{amsmath}
\usepackage{graphicx}
\usepackage{dcolumn}
\usepackage{float}
\begin{document}

\title{Realistic shell-model calculations for isotopic chains ``north-east'' of
  $^{48}$Ca in the ($N,Z$) plane}

\author{L. Coraggio}
\affiliation{Istituto Nazionale di Fisica Nucleare, \\
Complesso Universitario di Monte  S. Angelo, Via Cintia - I-80126 Napoli,
Italy}
\author{A. Covello}
\affiliation{Istituto Nazionale di Fisica Nucleare, \\
Complesso Universitario di Monte  S. Angelo, Via Cintia - I-80126 Napoli,
Italy}
\author{A. Gargano}
\affiliation{Istituto Nazionale di Fisica Nucleare, \\
Complesso Universitario di Monte  S. Angelo, Via Cintia - I-80126 Napoli,
Italy}
\author{N. Itaco}
\affiliation{Istituto Nazionale di Fisica Nucleare, \\
Complesso Universitario di Monte  S. Angelo, Via Cintia - I-80126 Napoli,
Italy}
\affiliation{Dipartimento di Fisica, Universit\`a
di Napoli Federico II, \\
Complesso Universitario di Monte  S. Angelo, Via Cintia - I-80126 Napoli,
Italy}

\date{\today}

\begin{abstract}
We perform realistic shell-model calculations for nuclei with
valence nucleons outside $^{48}$Ca, employing two different model
spaces.
The matrix elements of the effective two-body interaction and 
electromagnetic multipole operators have been calculated within
the framework of the many-body perturbation theory, starting from a
low-momentum potential derived from the high-precision CD-Bonn free
nucleon-nucleon potential.
The role played by the neutron orbital $1d_{5/2}$ has been
investigated by comparing experimental data on yrast quadrupole
excitations of isotopic chains north-east of $^{48}$Ca with the
results of calculations including or not including this
single-particle state in the model space.
\end{abstract}


\maketitle

\section{Introduction}
An interesting aspect of the physics of nuclei approaching the neutron
drip line is the evolution of their shell structure.
This topic is currently investigated in different mass regions, and
the modern advances in the detection techniques and new experimental
devices provide data that drive to a better understanding of the
microscopic mechanism underlying modifications of the ``magic numbers''.

In this context, in the last decade it has been recognized the key
role played by the tensor component of the residual two-body
interaction between spin-orbit partner single-particle states
\cite{Otsuka01b,Otsuka05,Otsuka10b}.
This may give rise to a breaking of shell closures leading to the
possible appearance of the so-called "island of inversion", namely a
region  of nuclei where a rapid development of collectivity  is
observed. The best well-known example of this phenomenon  is given by
neutron-rich nuclei around $^{32}$Mg \cite{Brown10}.

The region of nuclei with valence protons outside doubly-closed
$^{48}$Ca may be considered an interesting laboratory to study the
shell evolution when adding neutrons, since there are long isotopic
chains, such as those of iron and nickel isotopes, whose exoticity
reaches an $N/Z$ value of 1.79 in $^{78}$Ni.

As a matter of fact, it can be observed that the shell closure at
$N=40$ in $^{68}$Ni, corresponding to the filling of the neutron $fp$
orbitals and of the proton $f_{7/2}$ orbital, rapidly disappears
when removing protons from $f_{7/2}$, as testified by the behavior of
the experimental excitation energies of the yrast $J^{\pi}=2^+$ states
in iron and chromium neutron-rich isotopes.

For this isotopic chains several experimental studies (see for example
\cite{Sorlin03,Adrich08,Ljungvall10,Rother11,Baugher12}) have found
out that the disappeareance of the $N=40$ shell closure comes along
with the onset of a collective behavior, as indicated for instance by
the observation of a rapid increase of the $E_x(4^+_1)/E_x(2^+_1)$
ratio in $^{60-64}$Cr \cite{Aoi09,Gade10}.
This has been related to the correlations between the
quadrupole-partner neutron orbitals $0g_{9/2}$ and $1d_{5/2}$
\cite{Zuker95,Caurier02,Lenzi10}, so, within a shell-model description
of these nuclei, the inclusion of the neutron $1d_{5/2}$ orbital
should be needed.

It is worth pointing out that there is a general belief
\cite{Smirnova12} that shell-model effective interactions derived from
realistic nucleon-nucleon ($NN$) potentials are defective in their
monopole component, so that they are not able to provide a good
description of the evolution of spectroscopic properties along the
isotopic chains, unless including contributions from three-nucleon
forces \cite{Schwenk06,Holt13}.

On these grounds, we have found it challenging to perform realistic
shell-model calculations \cite{Coraggio09a} for some isotopic chains
north-east of $^{48}$Ca, starting from the high-precision CD-Bonn $NN$
potential renormalized by way of the so-called $V_{\rm low-k}$
approach \cite{Bogner01,Bogner02}.
In particular, in order to explore the microscopic processes leading
to the disappearance of the $N=40$ shell closure and the onset of
collectivity, we have derived two effective shell-model interactions
for two different model spaces.
The first one, which is spanned by the proton $0f_{7/2}$ and $1p_{3/2}$
orbitals and by the neutron $1p_{3/2}$, $1p_{1/2}$, $0f_{5/2}$,
$0g_{9/2}$ orbitals, is able, as we show in the following, to
reproduce the dropping of the $N=40$ magic number in iron and chromium
isotopes as well as many other spectroscopic properties of nuclei
north-east of $^{48}$Ca.
The second one is spanned by the same orbitals plus the neutron
$1d_{5/2}$.
The calculations within the latter model space are able to reproduce
the onset of collectivity in heavy iron and chromium isotopes, at the
same time preserving the results obtained with the first model space.

In the following section, we outline the perturbative approach to the
derivation of our shell-model hamiltonians and effective charges of
the electric quadrupole operators. 
In Section III, we present the results of our calculations for
calcium, titanium, chromium, iron, and nickel isotopes.
Concluding remarks are given in the last section.
In the Appendix, we report the calculated two-body matrix elements
(TBME) of the effective shell-model interactions, the employed
single-particle (SP) energies, and the effective charges of the $E2$
operator.

\section{Outline of calculations}
Within the framework of the shell model, an auxiliary one-body
potential $U$ is introduced in order to break up the hamiltonian for a
system of $A$ nucleons as the sum of a one-body term $H_0$, which
describes the independent motion of the nucleons, and a residual
interaction $H_1$:

\begin{eqnarray}
H=\sum_{i=1}^{A} \frac{p_i^2}{2m} + \sum_{i<j=1}^{A} V_{ij}^{NN} = T + V =
(T+U) +\nonumber \\
+(V-U)= H_{0}+H_{1}~~.~~~~~~~~~~~~~~~~~~~~~~~~~
\label{smham}
\end{eqnarray}

Once $H_0$ has been introduced, it is possible to define a reduced
model space in terms of a finite subset of $H_0$'s eigenvectors. 
In this space, an effective hamiltonian $H_{\rm eff}$ may be
constructed and the diagonalization of the many-body hamiltonian
(\ref{smham}) in an infinite Hilbert space is then reduced to the
solution of an eigenvalue problem in a finite space.

As mentioned in the Introduction, we derive effective shell-model
hamiltonians $H_{\rm eff}$ for two model spaces outside $^{48}$Ca
core.
The first one (I) is spanned by the two single-proton levels
$0f_{7/2}$, $1p_{3/2}$ and the four single-neutron levels $1p_{3/2},
1p_{1/2}$, $0f_{5/2}$, $0g_{9/2}$, while the second one (II) includes
also the neutron $1d_{5/2}$ orbital.

To this end, the following approach has been employed: first we have
derived a $H_{\rm eff}$ defined in a very large model space outside
the $^{40}$Ca closed core and spanned by six proton and neutron $pfgd$
orbitals.
Then, we derive from this $H_{\rm eff}$ two new effective
hamiltonians, defined in the smaller model spaces (I) and (II).
The above double-step procedure ensures that, when the $1d_{5/2}$
orbital is irrelevant to describe the physics of certain nuclear
states, the eigenvalues obtained using model space (I) or (II) are the
same or at least very close to each other.
 
The derivation of both $H_{\rm eff}$s has been done in the framework of
the time-dependent perturbation theory by way of the Kuo-Lee-Ratcliff
(KLR) folded-diagram expansion \cite{Kuo90}.
More precisely, we first renormalize the high-momentum repulsive 
components of the CD-Bonn $NN$ potential by way of the $V_{\rm low-k}$
approach \cite{Bogner01,Bogner02} with a momentum cutoff $\Lambda=2.6$
fm$^{-1}$, which provides a smooth potential preserving exactly the
on-shell properties of the original $NN$ potential.
Next, we express $H_{\rm eff}$ in terms of the vertex function
$\hat{Q}$-box, which is composed of irreducible valence-linked
diagrams \cite{Kuo71,Kuo81}.
In the derivation of the $\hat{Q}$-box we have included one- and
two-body Goldstone diagrams through third order in $H_1$
\cite{Coraggio12a}.
Since calculations beyond the third order in perturbation theory are
computationally prohibitive, we have calculated the Pad\'e approximant
$[2|1]$ \cite{Baker70,Ayoub79} of the $\hat{Q}$-box, so as to obtain a
value to which the perturbation series should converge, as suggested
in \cite{Hoffmann76}.
The folded-diagram series is then summed up to all orders using the
Lee-Suzuki iteration method \cite{Suzuki80}.

The hamiltonian $H_{\rm eff}$ contains one-body contributions, whose
collection is the so-called $\hat{S}$-box \cite{Shurpin83}. 
In realistic shell-model calculations a subtraction procedure is
commonly used, so as to retain only the TBME of $H_{\rm eff}$ ($V_{\rm
  eff}$), while the SP energies are taken from experiment.
This approach enables to take into account implicitly the effects of
three-body forces on the SP energies.

In this work, the proton SP energy spacing is taken from the
experimental one between the yrast $\frac{3}{2}^-$ state and the
$\frac{7}{2}^-$ ground state in $^{49}$Sc \cite{nndc}, which
are mainly of single-particle nature \cite{Fortier80}, so that
$\epsilon_{p_{3/2}}-\epsilon_{f_{7/2}}=3.1$ MeV.
Similarly, the two neutron $\epsilon_{p_{1/2}}-\epsilon_{p_{3/2}}$,
$\epsilon_{f_{5/2}}-\epsilon_{p_{3/2}}$ spacings are taken from the
experimental spectrum of $^{49}$Ca, considering the yrast
$\frac{1}{2}^-$ and the yrare $\frac{5}{2}^-$ states \cite{nndc}, that
have the largest observed single-particle components \cite{Cottle08}.
As regards the neutron $0g_{9/2}$ state, at present there is no
experimental indication of the existence of such a state in
$^{49}$Ca, therefore we have fixed its energy so as to reproduce the
experimental odd-even mass difference around $^{68}$Ni \cite{Audi03}.
All the observed $\frac{5}{2}^+$ states in $^{49}$Ca do not exhibit a
significant SP component, so we have chosen a value of the $1d_{5/2}$
energy which gives, when using model space (II), an odd-even mass
difference around $^{80}$Zn in a reasonable agreement with the
estimated value \cite{Audi03}.
In summary, our adopted neutron SP energy spacings for model space (I)
are $\epsilon_{p_{1/2}}-\epsilon_{p_{3/2}}=2.0$ MeV,
$\epsilon_{f_{5/2}}-\epsilon_{p_{3/2}}=4.0$ MeV,
$\epsilon_{g_{9/2}}-\epsilon_{p_{3/2}}=4.1$ MeV, and for model space
(II) $\epsilon_{g_{9/2}}-\epsilon_{p_{3/2}}=3.8$ MeV and
$\epsilon_{d_{5/2}}-\epsilon_{p_{3/2}}=8.0$ MeV.

The TBME and SP energies relative to $^{48}$Ca used in the present
calculation can be found in the Appendix, together with the effective
charges of the electric quadrupole operator that has been derived
consistently with the $H_{\rm eff}$.
It should be pointed out that for protons the Coulomb force has been
explicitly added to $V_{\rm low-k}$ before constructing $V_{\rm eff}$.

\section{Results}
In this section, we report the results of our shell-model calculations
for calcium, titanium, chromium, iron and nickel isotopes with mass
number $A > 48$, and compare them with the available experimental data.
As mentioned in the Introduction, our attention is focused on the
insurgence of collective behavior or shell-closure properties when
adding valence neutrons.

The calculations have been performed by using the Oslo shell-model
code \cite{EngelandSMC}.

\subsection{Calcium isotopes} \label{ca}

\begin{figure}[H]
\begin{center}
\includegraphics[scale=0.33,angle=0]{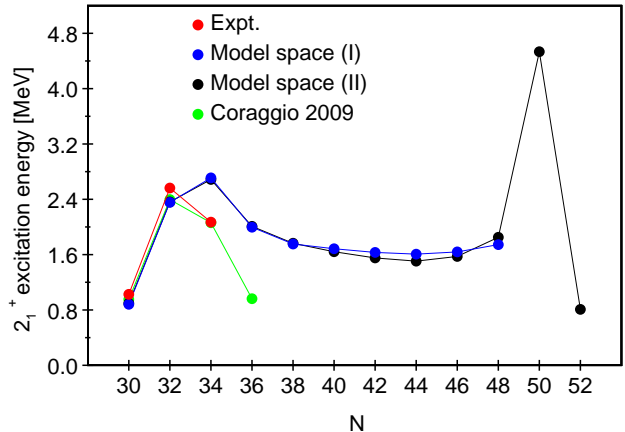}
\caption{(Color online) Experimental \cite{nndc,Steppenbeck13} and
  calculated excitation energies of the yrast $J^{\pi}=2^+$ states for
  calcium isotopes.}
\label{J02p_Ca}
\end{center}
\end{figure}
 A major topic in nuclear physics is the understanding of the shell
evolution towards the neutron dripline, and in this context the
study of calcium isotopes provides an interesting laboratory
\cite{Gridnev06,Capelli09,Duguet12,Hagen12,Holt13}.

An extensive study of the heavy calcium isotopes have been reported in
Ref. \cite{Coraggio09c}, where we have performed a shell-model
calculation using $^{40}$Ca as an inert core and the full $fp$ shell
as model space.
As in the present paper, the effective interaction has been
obtained from a $V_{\rm low-k}$ derived from the CD-Bonn potential
with a cutoff momentum equal to $2.6~{\rm fm}^{-1}$.
The results we have obtained are in excellent agreement with the
experimental data, in particular we have successfully predicted the
excitation energy of the yrast $J=2^+$ in $^{54}$Ca, that only
recently has been measured at RIKEN \cite{Steppenbeck13}.

In Fig. \ref{J02p_Ca} are reported the excitation energies of the
yrast $2^+$ states calculated with model spaces (I) and (II) as a
function of the neutron number up to $N=52$ and compared with the
available experimental data and the results obtained by our previous
calculations with the full $fp$-shell model space.

It can be seen that the available observed energies are well 
reproduced for $N=30,32$, the three calculations providing almost
the same values.
At $N=34,36$ the results with model spaces (I) and (II) differ from
those of Ref. \cite{Coraggio09c}, and are almost indistinguishable up
to $N=48$.
We have verified that to reproduce the observed drop in energy at
$N=34$ it is necessary, as in Ref. \cite{Coraggio09c}, to include
explicitly the neutron $0f_{7/2}$ orbital into the model space.
On the other hand, the different results obtained at $N=36$ trace back
to the presence in model spaces (I) and (II) of the $0g_{9/2}$ orbital
that starts to fill just at $N=36$.
As a matter of fact, we have verified that, when blocking out this
orbital, the calculated excitation energy of the yrast $2^+$ state
drops abruptly down at 0.7 MeV.

This suggests that to optimize the description of the calcium isotopic
chain one has to adopt a model space spanned by the full $fp$ shell
plus the $0g_{9/2}$ orbital.
\begin{figure}[H]
\begin{center}
\includegraphics[scale=0.33,angle=0]{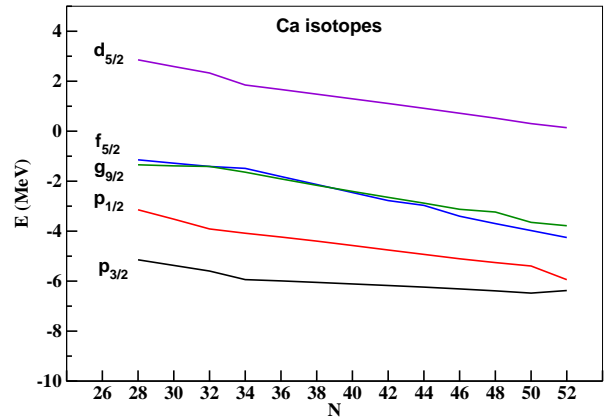}
\caption{(Color online) Calculated neutron effective single-particle
  energies for calcium isotopes from $N=30$ to 52.}
\label{ESPE_Ca}
\end{center}
\end{figure}
 From the inspection of Fig. \ref{J02p_Ca}, it can be seen that no shell
closure shows up at $N=40$, since the $0g_{9/2}$ SP energy is very
close to the $pf$ ones, while the large energy gap between $0g_{9/2}$
and $1d_{5/2}$ orbitals is responsible for a strong shell closure at
$N=50$ and that according to our calculations $^{70}$Ca should be the
last bound calcium isotope.
This picture is confirmed when looking at the behavior of the calcium
effective single-particle energies (ESPE), as reported in
Fig. \ref{ESPE_Ca}.

\subsection{Titanium isotopes} \label{ti}

\begin{figure}[H]
\begin{center}
\includegraphics[scale=0.33,angle=0]{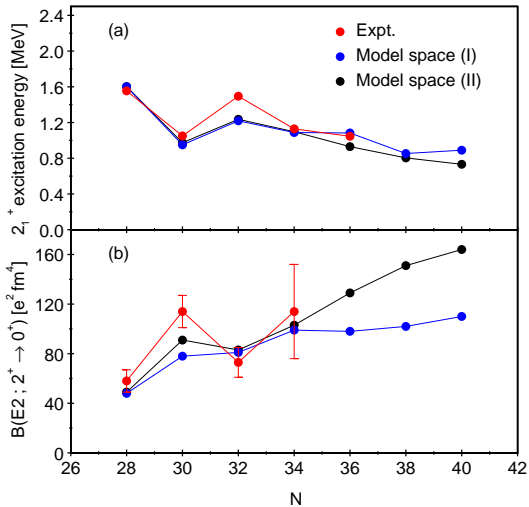}
\caption{(Color online) (a) Experimental \cite{nndc,Suzuki13} and
  calculated excitation energies of the yrast $J^{\pi}=2^+$ states and
  (b) $B(E2;2^+_1 \rightarrow 0^+_1)$  transition rates for titanium
  isotopes.}
\label{J02p_Ti}
\end{center}
\end{figure}

The calculated excitation energies of the yrast $2^+$ states and the
$B(E2;2^+_1 \rightarrow 0^+_1)$ transition rates for titanium isotopes
are reported in Fig. \ref{J02p_Ti} up to $N=40$.
It can be seen that the observed excitation energies and $B(E2)$s are
well reproduced by both calculations.

The decrease of the experimental $2^+_1$ excitation energy in $^{56}$Ti
is linked to the lack of a $N=34$ shell closure, as we predicted in
\cite{Coraggio09c} and confirmed by the recent observation of the
excitation energy of $J^{\pi}=2^+_1$ state in $^{54}$Ca
\cite{Steppenbeck13}.
From the inspection of Fig. \ref{J02p_Ti}, it can be observed that the
calculated $2^+_1$ excitation energies are almost insensitive to
the inclusion of the $1d_{5/2}$ orbital. 
This does not happen with the $B(E2)$s, whose calculated values begin
to differ substantially starting from $N=36$ on.
In particular, using model space (I) the increase of the number of
valence neutrons corresponds to a rise of the proton $0f_{7/2}$
occupation number and to the decrease of the proton polarization, at
variance with model space (II) where the occupation of this
orbital is almost constant respect to the neutron number $N$ (see
Table \ref{occupations}).
This comes with a significant larger occupation of the neutron
$0g_{9/2}$ orbital with model space (II) with respect to model space
(I), as well as an increase of the neutron $1d_{5/2}$-orbital
occupation that provides a constant rise of the corresponding $B(E2)$s
up to $N=40$.

\subsection{Chromium isotopes} \label{cr}

\begin{figure}[H]
\begin{center}
\includegraphics[scale=0.33,angle=0]{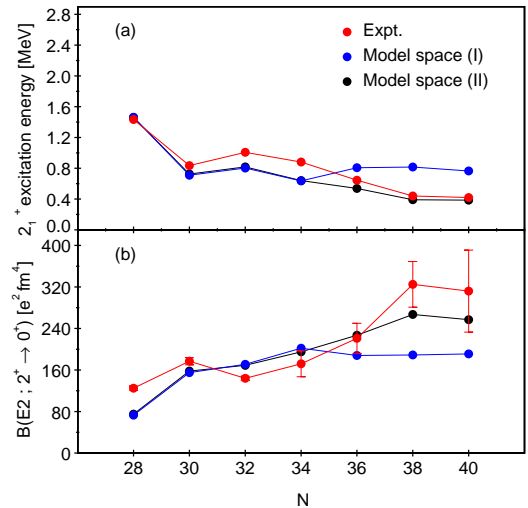}
\caption{(Color online) Same as in Fig. \ref{J02p_Ti}, but for
  chromium isotopes. Experimental data are taken from \cite{nndc,xundl}}
\label{J02p_Cr}
\end{center}
\end{figure}

The excitation energies of the yrast $2^+$ states and $B(E2;2^+_1
\rightarrow 0^+_1)$ transition rates for the chromium isotopes are
reported in Fig. \ref{J02p_Cr} up to $N=40$.
It is worth mentioning that the observed lowering of the yrast $2^+$
states starting from $N=34$ is interpreted as a signature of the
development of a pronounced collectivity towards $N=40$ \cite{Gade10}.
The interpretation of such a collective behavior may be framed within
the quasi-SU(3) approximate symmetry owing to the interplay between
the quadrupole-quadrupole component of the residual interaction and
the central field in the sub-space spanned by the lowest $\Delta j$=2
orbitals of a major shell \cite{Zuker95}.
In this connection, the need to include the neutron $1d_{5/2}$ orbital
in a shell-model calculation, in order to produce the onset of
collectivity around $N=40$ in chromium isotopes, has been evidenced by
Caurier, Nowacki, and Poves in Ref. \cite{Caurier02}.
\begin{figure}[H]
\begin{center}
\includegraphics[scale=0.33,angle=0]{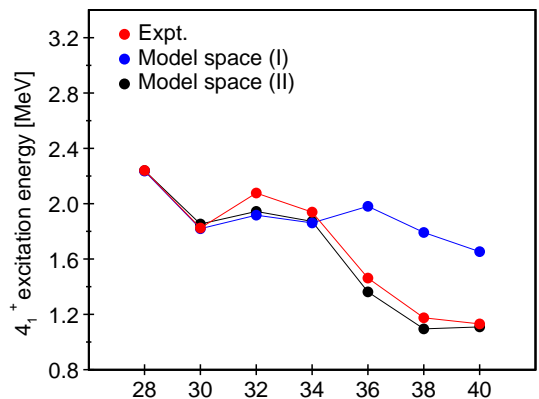}
\caption{(Color online) Experimental \cite{nndc} and
  calculated excitation energies of the yrast $J^{\pi}=4^+$. See text
  for details.}
\label{J04p_Cr}
\end{center}
\end{figure}
 Also our results support the crucial role played by this orbital.
In fact, from the inspection of Fig. \ref{J02p_Cr} it can be observed
that using model space (I) from $^{58}$Cr to $^{60}$Cr the calculated
$J^{\pi}=2^+_1$ states start to rise in excitation energy, while the
experimental behavior does the opposite.
Actually, employing the model space (II) there is a decrease of the
$2^+$ excitation energy from $N=34$ to $N=36$.
From the inspection of Table \ref{occupations}, we see that the
appearance of the differencies between model-space (I) and (II)
results come together with an abrupt increase of the occupation
number of the neutron $1d_{5/2},~0g_{9/2}$ orbitals and a depletion
of the proton $0f_{7/2}$ one.
This has to be ascribed to a reduced neutron gap between $0g_{9/2} -
0f_{5/2}$ (about 2.5 MeV) effective single-particle energies (ESPE) 
with respect to the one in Ni isotopes (about 4.0 MeV), where a shell
closure is found at $N=40$ (see Sec. \ref{conclusions}), as well as to
the quadrupole-quadrupole component of the effective interaction
between the $1d_{5/2}$ and the $0g_{9/2}$ orbitals.

\begin{table*}
\caption{Occupation numbers of proton $0f_{7/2}$ and neutron
  $0g_{9/2},1d_{5/2}$ of $^{56-62}$Ti, $^{58-64}$Cr, $^{60-66}$Fe, and
  $^{62-68}$Ni calculated with model spaces (I) and (II) (see
  text for details).}
\label{occupations}
\smallskip
\begin{tabular*}{\textwidth}{@{\extracolsep{\fill}}cccccccccc}
\hline
\hline
\noalign{\smallskip}
  & & \multicolumn{2}{c}{N=34} & \multicolumn{2}{c}{N=36} &
\multicolumn{2}{c}{N=38} & \multicolumn{2}{c}{N=40} \\
  & & (I) & (II) & (I) & (II) & (I) & (II) & (I) & (II)  \\
\hline
\noalign{\smallskip}
   &$\pi 0f_{7/2}$ & 1.71 & 1.71 & 1.84 & 1.72 & 1.87 & 1.71 & 1.87 &
1.69 \\
Ti &$\nu 0g_{9/2}$ & 0.24 & 0.31 & 0.95 & 1.45 & 2.15 & 2.61 & 3.23 &
3.72 \\
   &$\nu 1d_{5/2}$ &   ~  & 0.05 &   ~  & 0.17 &   ~  & 0.27 &   ~  &
0.36 \\
\hline
\noalign{\smallskip}
   &$\pi 0f_{7/2}$ & 3.37 & 3.37 & 3.54 & 3.33 & 3.60 & 3.26 & 3.64 &
3.31 \\
Cr &$\nu 0g_{9/2}$ & 0.14 & 0.18 & 0.66 & 1.48 & 1.68 & 2.98 & 2.88 &
3.73 \\
   &$\nu 1d_{5/2}$ &   ~  & 0.04 &   ~  & 0.26 &   ~  & 0.53 &   ~  &
0.57 \\
\hline
\noalign{\smallskip}
  &$\pi 0f_{7/2}$ & 5.36 & 5.35 & 5.35 & 5.34 & 5.49 & 5.31 & 5.61 &
5.31 \\
Fe &$\nu 0g_{9/2}$ & 0.10 & 0.12 & 0.27 & 0.37 & 0.89 & 1.51 & 1.75 &
2.81 \\
   &$\nu 1d_{5/2}$ &   ~  & 0.03 &   ~  & 0.06 &   ~  & 0.17 &   ~  &
0.33 \\
\hline
\noalign{\smallskip}
   &$\pi 0f_{7/2}$ & 6.77 & 6.76 & 6.92 & 6.91 & 7.33 & 7.32 & 7.89 &
7.85 \\
Ni &$\nu 0g_{9/2}$ & 0.09 & 0.10 & 0.16 & 0.17 & 0.25 & 0.26 & 0.42 &
0.47 \\
   &$\nu 1d_{5/2}$ &   ~  & 0.02 &   ~  & 0.03 &   ~  & 0.04 &   ~  &
0.05 \\
\hline
\hline
\noalign{\smallskip}
\end{tabular*}
\end{table*}

 Another feature that reveals the onset of the collectivity in
chromium isotopes is the rise of the experimental $B(E2;2^+_1
\rightarrow 0^+_1)$ transition rates from $N=34$ to $N=38$.
This can be reproduced only by calculations with model space (II),
as can be seen in Fig. \ref{J02p_Cr}, where the experimental data are
compared with our shell-model results with both model spaces.

In Fig. \ref{J04p_Cr} the calculated and experimental excitation
energies of the yrast $4^+$ states are also reported, evidencing that
only with model space (II) it is possible to reproduce the observed
increase of the $E_x(4^+_1)/E_x(2^+_1)$ ratio from $^{60}$Cr to $^{62}$Cr.

\subsection{Iron isotopes} \label{fe}
\begin{figure}[H]
\begin{center}
\includegraphics[scale=0.33,angle=0]{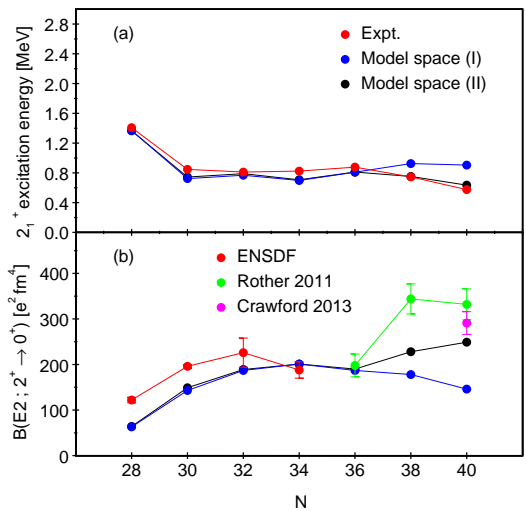}
\caption{(Color online) Same as in Fig. \ref{J02p_Ti}, but for
  iron isotopes. Experimental data are taken from ENSDF:\cite{nndc},
  Rother 2011:\cite{Rother11}, and Crawford 2013:\cite{Crawford13}.}
\label{J02p_Fe}
\end{center}
\end{figure}
In Fig. \ref{J02p_Fe} the $2^+_1$ excitation energies and
$B(E2;2^+_1 \rightarrow 0^+_1)$ transition rates of even iron
isotopes are shown as a function of the neutron number up to $N=40$.
It should be mentioned that experimental data are available for
$^{68}$Fe too \cite{Crawford13}, but they are not reported since for
this nucleus calculations performed with model space (II) need a CPU
time which exceeds our present computational resources.

A relevant feature that can be inferred from Fig. \ref{J02p_Fe} is the
abrupt enhancement of the $B(E2;2^+_1 \rightarrow 0^+_1)$ in $^{64}$Fe
\cite{Ljungvall10,Rother11,Crawford13} that, as in chromium isotopes,
is the evidence of the onset of a quadrupole collectivity.
This is at variance with the results of calculations with model space
(I), that show a decrease of the $B(E2;2^+_1 \rightarrow 0^+_1)$ from
$^{62}$Fe to $^{64}$Fe.
Actually, the onset of the above mentioned quadrupole collectivity is
reproduced using model space (II), and, as in Ti and Cr isotopes, comes
with a rise of the occupation number of the neutron $0g_{9/2}$ orbital
(see Table \ref{occupations}).
It has to be pointed out, however, that the experimental datum in
$^{64}$Fe \cite{Rother11} is underestimated.

\subsection{Nickel isotopes} \label{ni}
\begin{figure}[H]
\begin{center}
\includegraphics[scale=0.35,angle=0]{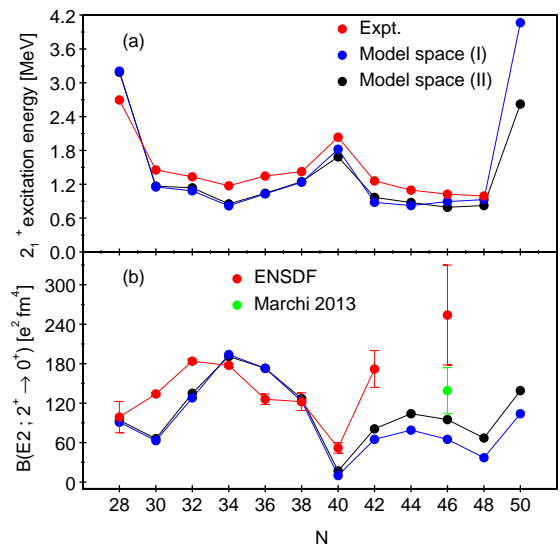}
\caption{(Color online) Same as in Fig. \ref{J02p_Ti}, but for
  nickel isotopes. Experimental data are taken from ENSDF:\cite{nndc}
  and Marchi 2013:\cite{Marchi13}.}
\label{J02p_Ni}
\end{center}
\end{figure}
The nickel isotopes represent a very interesting laboratory to
study the effects of the $NN$ interaction on the shell evolution, when
increasing the number of valence neutrons.
In fact, this is a very long isotopic chain with available
spectroscopic data from $N=24$ up to $N=48$, and several efforts are
at present devoted to get as much experimental information as
possible on $^{78}$Ni, which is a possible waiting point for
$r-$process nucleosyntesis \cite{Hosmer05}.
In Fig. \ref{J02p_Ni} they are reported the experimental data and the
results of our shell-model calculations with model spaces (I) and (II)
for the $2^+_1$ excitation energy and the $B(E2,2^+_1\rightarrow 0^+_1)$
systematics from $N=28$ to $N=50$.

\begin{figure}[H]
\begin{center}
\includegraphics[scale=0.33,angle=0]{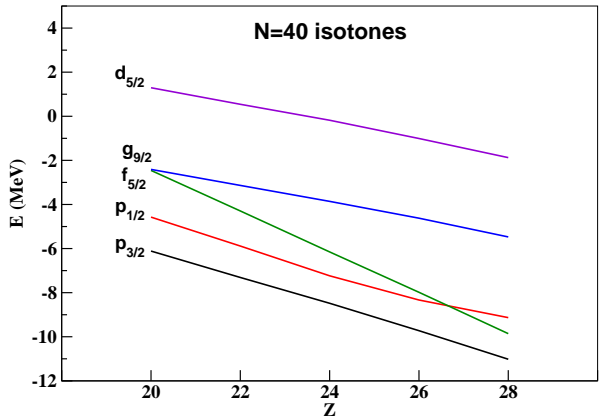}
\caption{(Color online) Calculated neutron effective single-particle
  energies for Ti, Cr, Fe, and Ni isotopes at $N=40$.}
\label{ESPE}
\end{center}
\end{figure}

As already mentioned in the Introduction, it can be seen that the
filling of the proton $0f_{7/2}$ orbital, corresponding to the
well-known $Z=28$ closure, produces the disappearance of the
collectivity towards $N=40$ observed in chromium and iron isotopes.
The peak of the yrast $2^+$ excitation energy and the fall of the
corresponding $B(E2)$ in $^{68}$Ni is a clear manifestation of the
shell closure at $N=40$, which is added to the ones at $N=28$ and
$N=50$ in this long isotopic chain.
As a matter of fact, it turns out that our calculated neutron ESPE
provide a large gap between the $0g_{9/2}$ and $fp$ orbitals (see
Fig. \ref{ESPE}).
This implies a minor role played by the correlations between the
quadrupole partners $0g_{9/2},~1d_{5/2}$, as supported by the fact
that the calculations with model space (I) and (II) give similar
occupation numbers (see Table \ref{occupations}), as well as the
same behavior for the $E_x(2^+_1)$ and the electric-quadrupole
transition rates (see Fig. \ref{J02p_Ni}).
\begin{figure}[H]
\begin{center}
\includegraphics[scale=0.33,angle=0]{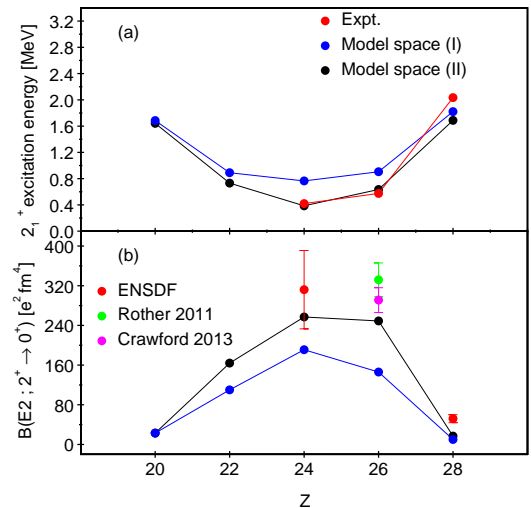}
\caption{(Color online) (a) Experimental (ENSDF:\cite{nndc},
  Rother 2011:\cite{Rother11}, Crawford 2013:\cite{Crawford13}) and
  calculated excitation energies of the yrast $J^{\pi}=2^+$ states and
  (b) $B(E2;2^+_1 \rightarrow 0^+_1)$  transition rates for the $N=40$
  isotones.}
\label{J02p_N40}
\end{center}
\end{figure}
It is interesting to note that recently the $B(E2)$ in $^{74}$Ni has
been measured in a Coulomb excitation experiment \cite{Marchi13},
the obtained value being significantly smaller than the one measured
indirectly via a $(p,p')$ inhelastic diffusion \cite{Aoi10}.
Our values of the $B(E2)$s for $^{74}$Ni are close to the recently
measured one.
They,  however, do not  reproduce the decrease in the $B(E2)$ from
$N=42$ to $N=46$ since the $^{70}$Ni $B(E2)$ is significantly
underestimated.
More precisely, we predict an increase of the $B(E2)$ from $N=40$ to
42 but not as large as the observed one.
This may be traced to the fact that our proton model space leaves out
the $0f_{5/2}$ orbital, which has a relevant role  beyond $N=40$ as
evidenced by the behavior of the yrast $\frac{5}{2}^-$ states in
odd-mass Cu isotopes \cite{nndc}.
As a matter of fact, for Ni isotopes with $N>40$ the attraction due to
the tensor force between the neutron $0g_{9/2}$ and the proton
$0f_{5/2}$ orbitals may give rise to a lowering of the latter leading
to a strong core polarization with a consequent increase of the
$B(E2)$s.
It is worth mentioning, however, that shell-model calculations that
include explicitly the proton $0f_{5/2}$ orbital show the same
behavior of the $B(E2)$ as ours \cite{Shimizu12}.

Finally, it has to be observed that, because of the magicity at
$N=50$, the yrast $J=2^+$ state in $^{78}$Ni is built up breaking a
couple of protons in the filled $0f_{7/2}$ orbital and promoting
one proton into the $1p_{3/2}$ orbital. 
This mechanism is then responsible of the increase of the theoretical
$B(E2;2^+\rightarrow 0^+$) with respect to neighboring nuclei.

\section{Concluding remarks} \label{conclusions}
We have presented here the results of an extensive shell-model study
of nuclei north-east of $^{48}$Ca within the framework of a
microscopic approach.
This means that, starting from a $V_{\rm low-k}$ derived from the
CD-Bonn $NN$ potential, effective two-body interactions and effective
electromagnetic multipole operators have been calculated by way of
perturbation theory and then employed in shell-model calculations.

As mentioned in the Introduction, this region is currently the subject
of many experimental and theoretical investigations, especially because
of the behavior of the quadrupole collectivity at $N=40$ versus the
atomic number, as can be inferred from the inspection of
Fig. \ref{J02p_N40}.

In this context, we have focused attention on the role played by the
neutron $1d_{5/2}$ orbital in the spectroscopy of the nuclei under
investigation, using two different model spaces which differ only by
the inclusion of this SP state.

Our results are in a better agreement with experiment when using the
larger model space, in line with other large-scale shell-model
calculations \cite{Kaneko08,Lenzi10,Poves12,Shimizu12} where the TBME
were determined empirically.
This confirms the ability of realistic shell-model calculations to
provide a reliable microscopic description of the shell evolution
along isotopic chains, even in presence of strong collective features.
In fact, results with model spaces (I) and (II) are similar only for
Ni isotopes, evidencing the relevance of including the neutron
$1d_{5/2}$ orbital to describe the collective properties of Ti, Cr,
and Fe isotopic chains.
The role played by $1d_{5/2}$ may be explained in terms of correlations
between this orbital and its quadrupole partner $0g_{9/2}$, which
comes into play because of the reduction of the $0f_{5/2}-0g_{9/2}$
gap.

To clarify the above considerations, our results can be resumed
reporting in Figs. \ref{ESPE},\ref{J02p_N40} the ESPE, as well as the
energies of the yrast $2+$ states and the $B(E2;2^+ \rightarrow 0+$s,
at $N=40$ as a function of $Z$, respectively.
From the inspection of Fig. \ref{J02p_N40}, it can be clearly seen
that the nature of the yrast $2^+$ states changes with the
filling of the proton $0f_{7/2}$ orbital, tuning the collectivity at
$N=40$.

It has to be stressed that our calculations employ only a two-body
 $V_{\rm low-k}$, derived from the high-precision CD-Bonn potential,
not taking into account any three-body forces.
Therefore, the good quality of the agreement between theory and
experiment seems to question the role of valence three-body forces in
the description of the spectroscopic properties of intermediate-mass
nuclei.

\bibliographystyle{apsrev}
\bibliography{biblio}

\newpage
\section*{Appendix}
\renewcommand{\thetable}{A.\Roman{table}}
\setcounter{table}{0}

\begin{table}[H]
\caption{Shell-model proton single-particle energies (in MeV)
  employed in present work (see text for details). The value
  of the $\epsilon_{f_{7/2}}$ is taken from the one-proton separation
  energy of $^{49}$Sc \cite{Audi03}.}
\begin{ruledtabular}
\begin{tabular}{cc}
$nlj$ & Model space (I) and (II) \\
\colrule
$0f_{7/2}$  & -9.627 \\
$1p_{3/2}$  & -6.527 \\
\end{tabular}
\end{ruledtabular}
\label{protonspetab}
\end{table}

\begin{table}[H]
\caption{Shell-model neutron single-particle energies (in MeV)
  employed in present work (see text for details). The value
  of the $\epsilon_{p_{3/2}}$ is taken from the one-neutron separation
  energy of $^{49}$Ca \cite{Audi03}.}
\begin{ruledtabular}
\begin{tabular}{ccc}
$nlj$ & Model space (I) & Model space (II) \\
\colrule
$1p_{3/2}$  & -5.157  & -5.157 \\
$1p_{1/2}$  & -3.157  & -3.157 \\
$0f_{5/2}$  & -1.157  & -1.157 \\
$0g_{9/2}$  & -1.057  & -1.357 \\
$1d_{5/2}$  &   ~     &  2.843 \\
\end{tabular}
\end{ruledtabular}
\label{neutronspetab}
\end{table}

\begin{table}[H]
\caption{Proton-proton, neutron-neutron, and proton-neutron matrix
  elements (in MeV) derived for calculations in model space (I). They
  are antisymmetrized, and normalized by a factor $1/ \sqrt{ (1 +
    \delta_{j_aj_b})(1 + \delta_{j_cj_d})}$.}
\begin{ruledtabular}
\begin{tabular}{cccc}
$n_a l_a j_a ~ n_b l_b j_b ~ n_c l_c j_c ~ n_d l_d j_d $ & $J$ & $T_z$
  &  TBME \\
\colrule
 $ 0f_{ 7/2}~ 0f_{ 7/2}~ 0f_{ 7/2}~ 0f_{ 7/2}$ &  0 &  1 & -2.021 \\
 $ 0f_{ 7/2}~ 0f_{ 7/2}~ 1p_{ 3/2}~ 1p_{ 3/2}$ &  0 &  1 & -1.219 \\
 $ 1p_{ 3/2}~ 1p_{ 3/2}~ 1p_{ 3/2}~ 1p_{ 3/2}$ &  0 &  1 & -0.058 \\
 $ 0f_{ 7/2}~ 0f_{ 7/2}~ 0f_{ 7/2}~ 0f_{ 7/2}$ &  2 &  1 & -0.414 \\
 $ 0f_{ 7/2}~ 0f_{ 7/2}~ 0f_{ 7/2}~ 1p_{ 3/2}$ &  2 &  1 & -0.704 \\
 $ 0f_{ 7/2}~ 0f_{ 7/2}~ 1p_{ 3/2}~ 1p_{ 3/2}$ &  2 &  1 & -0.355 \\
 $ 0f_{ 7/2}~ 1p_{ 3/2}~ 0f_{ 7/2}~ 1p_{ 3/2}$ &  2 &  1 & -0.547 \\
 $ 0f_{ 7/2}~ 1p_{ 3/2}~ 1p_{ 3/2}~ 1p_{ 3/2}$ &  2 &  1 & -0.356 \\
 $ 1p_{ 3/2}~ 1p_{ 3/2}~ 1p_{ 3/2}~ 1p_{ 3/2}$ &  2 &  1 &  0.108 \\
 $ 0f_{ 7/2}~ 1p_{ 3/2}~ 0f_{ 7/2}~ 1p_{ 3/2}$ &  3 &  1 &  0.329 \\
 $ 0f_{ 7/2}~ 0f_{ 7/2}~ 0f_{ 7/2}~ 0f_{ 7/2}$ &  4 &  1 &  0.133 \\
 $ 0f_{ 7/2}~ 0f_{ 7/2}~ 0f_{ 7/2}~ 1p_{ 3/2}$ &  4 &  1 & -0.408 \\
 $ 0f_{ 7/2}~ 1p_{ 3/2}~ 0f_{ 7/2}~ 1p_{ 3/2}$ &  4 &  1 &  0.059 \\
 $ 0f_{ 7/2}~ 1p_{ 3/2}~ 0f_{ 7/2}~ 1p_{ 3/2}$ &  5 &  1 &  0.497 \\
 $ 0f_{ 7/2}~ 0f_{ 7/2}~ 0f_{ 7/2}~ 0f_{ 7/2}$ &  6 &  1 &  0.406 \\
 $ 1p_{ 3/2}~ 1p_{ 3/2}~ 1p_{ 3/2}~ 1p_{ 3/2}$ &  0 & -1 & -0.748 \\
 $ 1p_{ 3/2}~ 1p_{ 3/2}~ 1p_{ 1/2}~ 1p_{ 1/2}$ &  0 & -1 & -0.903 \\
 $ 1p_{ 3/2}~ 1p_{ 3/2}~ 0f_{ 5/2}~ 0f_{ 5/2}$ &  0 & -1 & -0.712 \\
 $ 1p_{ 3/2}~ 1p_{ 3/2}~ 0g_{ 9/2}~ 0g_{ 9/2}$ &  0 & -1 &  1.171 \\
 $ 1p_{ 1/2}~ 1p_{ 1/2}~ 1p_{ 1/2}~ 1p_{ 1/2}$ &  0 & -1 & -0.312 \\
 $ 1p_{ 1/2}~ 1p_{ 1/2}~ 0f_{ 5/2}~ 0f_{ 5/2}$ &  0 & -1 & -0.478 \\
 $ 1p_{ 1/2}~ 1p_{ 1/2}~ 0g_{ 9/2}~ 0g_{ 9/2}$ &  0 & -1 &  0.884 \\
 $ 0f_{ 5/2}~ 0f_{ 5/2}~ 0f_{ 5/2}~ 0f_{ 5/2}$ &  0 & -1 & -0.988 \\
 $ 0f_{ 5/2}~ 0f_{ 5/2}~ 0g_{ 9/2}~ 0g_{ 9/2}$ &  0 & -1 &  1.577 \\
 $ 0g_{ 9/2}~ 0g_{ 9/2}~ 0g_{ 9/2}~ 0g_{ 9/2}$ &  0 & -1 & -1.179 \\
\end{tabular}
\end{ruledtabular}
\label{tbmeI}
\end{table}

\begin{table}[H]
\begin{ruledtabular}
\begin{tabular}{cccc}
\colrule
 $ 1p_{ 3/2}~ 1p_{ 1/2}~ 1p_{ 3/2}~ 1p_{ 1/2}$ &  1 & -1 &  0.167 \\
 $ 1p_{ 3/2}~ 1p_{ 1/2}~ 1p_{ 3/2}~ 0f_{ 5/2}$ &  1 & -1 & -0.054 \\
 $ 1p_{ 3/2}~ 0f_{ 5/2}~ 1p_{ 3/2}~ 0f_{ 5/2}$ &  1 & -1 &  0.046 \\
 $ 1p_{ 3/2}~ 1p_{ 3/2}~ 1p_{ 3/2}~ 1p_{ 3/2}$ &  2 & -1 & -0.241 \\
 $ 1p_{ 3/2}~ 1p_{ 3/2}~ 1p_{ 3/2}~ 1p_{ 1/2}$ &  2 & -1 & -0.394 \\
 $ 1p_{ 3/2}~ 1p_{ 3/2}~ 1p_{ 3/2}~ 0f_{ 5/2}$ &  2 & -1 & -0.162 \\
 $ 1p_{ 3/2}~ 1p_{ 3/2}~ 1p_{ 1/2}~ 0f_{ 5/2}$ &  2 & -1 & -0.179 \\
 $ 1p_{ 3/2}~ 1p_{ 3/2}~ 0f_{ 5/2}~ 0f_{ 5/2}$ &  2 & -1 & -0.178 \\
 $ 1p_{ 3/2}~ 1p_{ 3/2}~ 0g_{ 9/2}~ 0g_{ 9/2}$ &  2 & -1 &  0.441 \\
 $ 1p_{ 3/2}~ 1p_{ 1/2}~ 1p_{ 3/2}~ 1p_{ 1/2}$ &  2 & -1 & -0.448 \\
 $ 1p_{ 3/2}~ 1p_{ 1/2}~ 1p_{ 3/2}~ 0f_{ 5/2}$ &  2 & -1 & -0.176 \\
 $ 1p_{ 3/2}~ 1p_{ 1/2}~ 1p_{ 1/2}~ 0f_{ 5/2}$ &  2 & -1 & -0.407 \\
 $ 1p_{ 3/2}~ 1p_{ 1/2}~ 0f_{ 5/2}~ 0f_{ 5/2}$ &  2 & -1 & -0.292 \\
 $ 1p_{ 3/2}~ 1p_{ 1/2}~ 0g_{ 9/2}~ 0g_{ 9/2}$ &  2 & -1 &  0.308 \\
 $ 1p_{ 3/2}~ 0f_{ 5/2}~ 1p_{ 3/2}~ 0f_{ 5/2}$ &  2 & -1 &  0.071 \\
 $ 1p_{ 3/2}~ 0f_{ 5/2}~ 1p_{ 1/2}~ 0f_{ 5/2}$ &  2 & -1 & -0.280 \\
 $ 1p_{ 3/2}~ 0f_{ 5/2}~ 0f_{ 5/2}~ 0f_{ 5/2}$ &  2 & -1 & -0.103 \\
 $ 1p_{ 3/2}~ 0f_{ 5/2}~ 0g_{ 9/2}~ 0g_{ 9/2}$ &  2 & -1 &  0.494 \\
 $ 1p_{ 1/2}~ 0f_{ 5/2}~ 1p_{ 1/2}~ 0f_{ 5/2}$ &  2 & -1 & -0.297 \\
 $ 1p_{ 1/2}~ 0f_{ 5/2}~ 0f_{ 5/2}~ 0f_{ 5/2}$ &  2 & -1 & -0.282 \\
 $ 1p_{ 1/2}~ 0f_{ 5/2}~ 0g_{ 9/2}~ 0g_{ 9/2}$ &  2 & -1 &  0.572 \\
 $ 0f_{ 5/2}~ 0f_{ 5/2}~ 0f_{ 5/2}~ 0f_{ 5/2}$ &  2 & -1 & -0.405 \\
 $ 0f_{ 5/2}~ 0f_{ 5/2}~ 0g_{ 9/2}~ 0g_{ 9/2}$ &  2 & -1 &  0.239 \\
 $ 0g_{ 9/2}~ 0g_{ 9/2}~ 0g_{ 9/2}~ 0g_{ 9/2}$ &  2 & -1 & -0.772 \\
 $ 1p_{ 3/2}~ 0f_{ 5/2}~ 1p_{ 3/2}~ 0f_{ 5/2}$ &  3 & -1 &  0.125 \\
 $ 1p_{ 3/2}~ 0f_{ 5/2}~ 1p_{ 1/2}~ 0f_{ 5/2}$ &  3 & -1 & -0.005 \\
 $ 1p_{ 1/2}~ 0f_{ 5/2}~ 1p_{ 1/2}~ 0f_{ 5/2}$ &  3 & -1 &  0.169 \\
 $ 1p_{ 3/2}~ 0f_{ 5/2}~ 1p_{ 3/2}~ 0f_{ 5/2}$ &  4 & -1 & -0.336 \\
 $ 1p_{ 3/2}~ 0f_{ 5/2}~ 0f_{ 5/2}~ 0f_{ 5/2}$ &  4 & -1 & -0.247 \\
 $ 1p_{ 3/2}~ 0f_{ 5/2}~ 0g_{ 9/2}~ 0g_{ 9/2}$ &  4 & -1 &  0.493 \\
 $ 0f_{ 5/2}~ 0f_{ 5/2}~ 0f_{ 5/2}~ 0f_{ 5/2}$ &  4 & -1 & -0.087 \\
 $ 0f_{ 5/2}~ 0f_{ 5/2}~ 0g_{ 9/2}~ 0g_{ 9/2}$ &  4 & -1 &  0.115 \\
 $ 0g_{ 9/2}~ 0g_{ 9/2}~ 0g_{ 9/2}~ 0g_{ 9/2}$ &  4 & -1 & -0.282 \\
 $ 0g_{ 9/2}~ 0g_{ 9/2}~ 0g_{ 9/2}~ 0g_{ 9/2}$ &  6 & -1 & -0.108 \\
 $ 0g_{ 9/2}~ 0g_{ 9/2}~ 0g_{ 9/2}~ 0g_{ 9/2}$ &  8 & -1 & -0.001 \\
 $ 0f_{ 5/2}~ 0g_{ 9/2}~ 0f_{ 5/2}~ 0g_{ 9/2}$ &  2 & -1 & -0.404 \\
 $ 1p_{ 3/2}~ 0g_{ 9/2}~ 1p_{ 3/2}~ 0g_{ 9/2}$ &  3 & -1 & -0.603 \\
 $ 1p_{ 3/2}~ 0g_{ 9/2}~ 0f_{ 5/2}~ 0g_{ 9/2}$ &  3 & -1 &  0.227 \\
 $ 0f_{ 5/2}~ 0g_{ 9/2}~ 0f_{ 5/2}~ 0g_{ 9/2}$ &  3 & -1 & -0.211 \\
 $ 1p_{ 3/2}~ 0g_{ 9/2}~ 1p_{ 3/2}~ 0g_{ 9/2}$ &  4 & -1 &  0.039 \\
 $ 1p_{ 3/2}~ 0g_{ 9/2}~ 1p_{ 1/2}~ 0g_{ 9/2}$ &  4 & -1 & -0.105 \\
 $ 1p_{ 3/2}~ 0g_{ 9/2}~ 0f_{ 5/2}~ 0g_{ 9/2}$ &  4 & -1 &  0.157 \\
 $ 1p_{ 1/2}~ 0g_{ 9/2}~ 1p_{ 1/2}~ 0g_{ 9/2}$ &  4 & -1 &  0.116 \\
 $ 1p_{ 1/2}~ 0g_{ 9/2}~ 0f_{ 5/2}~ 0g_{ 9/2}$ &  4 & -1 & -0.007 \\
 $ 0f_{ 5/2}~ 0g_{ 9/2}~ 0f_{ 5/2}~ 0g_{ 9/2}$ &  4 & -1 &  0.109 \\
 $ 1p_{ 3/2}~ 0g_{ 9/2}~ 1p_{ 3/2}~ 0g_{ 9/2}$ &  5 & -1 & -0.099 \\
 $ 1p_{ 3/2}~ 0g_{ 9/2}~ 1p_{ 1/2}~ 0g_{ 9/2}$ &  5 & -1 &  0.361 \\
 $ 1p_{ 3/2}~ 0g_{ 9/2}~ 0f_{ 5/2}~ 0g_{ 9/2}$ &  5 & -1 &  0.158 \\
 $ 1p_{ 1/2}~ 0g_{ 9/2}~ 1p_{ 1/2}~ 0g_{ 9/2}$ &  5 & -1 & -0.335 \\
 $ 1p_{ 1/2}~ 0g_{ 9/2}~ 0f_{ 5/2}~ 0g_{ 9/2}$ &  5 & -1 & -0.252 \\
 $ 0f_{ 5/2}~ 0g_{ 9/2}~ 0f_{ 5/2}~ 0g_{ 9/2}$ &  5 & -1 & -0.065 \\
 $ 1p_{ 3/2}~ 0g_{ 9/2}~ 1p_{ 3/2}~ 0g_{ 9/2}$ &  6 & -1 &  0.260 \\
 $ 1p_{ 3/2}~ 0g_{ 9/2}~ 0f_{ 5/2}~ 0g_{ 9/2}$ &  6 & -1 &  0.176 \\
 $ 0f_{ 5/2}~ 0g_{ 9/2}~ 0f_{ 5/2}~ 0g_{ 9/2}$ &  6 & -1 &  0.156 \\
 $ 0f_{ 5/2}~ 0g_{ 9/2}~ 0f_{ 5/2}~ 0g_{ 9/2}$ &  7 & -1 & -0.730 \\
 $ 1p_{ 3/2}~ 1p_{ 3/2}~ 1p_{ 3/2}~ 1p_{ 3/2}$ &  0 &  0 & -2.071 \\
 $ 0f_{ 7/2}~ 0f_{ 5/2}~ 0f_{ 7/2}~ 0f_{ 5/2}$ &  1 &  0 & -1.994 \\
 $ 0f_{ 7/2}~ 0f_{ 5/2}~ 1p_{ 3/2}~ 1p_{ 3/2}$ &  1 &  0 &  0.798 \\
 $ 0f_{ 7/2}~ 0f_{ 5/2}~ 1p_{ 3/2}~ 1p_{ 1/2}$ &  1 &  0 & -0.970 \\
 $ 0f_{ 7/2}~ 0f_{ 5/2}~ 1p_{ 3/2}~ 0f_{ 5/2}$ &  1 &  0 & -0.684 \\
 $ 1p_{ 3/2}~ 1p_{ 3/2}~ 1p_{ 3/2}~ 1p_{ 3/2}$ &  1 &  0 & -0.388 \\
 $ 1p_{ 3/2}~ 1p_{ 3/2}~ 1p_{ 3/2}~ 1p_{ 1/2}$ &  1 &  0 &  0.617 \\
 $ 1p_{ 3/2}~ 1p_{ 3/2}~ 1p_{ 3/2}~ 0f_{ 5/2}$ &  1 &  0 & -0.018 \\
\end{tabular}
\end{ruledtabular}
\end{table}


\begin{table}[H]
\begin{ruledtabular}
\begin{tabular}{cccc}
\colrule
 $ 1p_{ 3/2}~ 1p_{ 1/2}~ 1p_{ 3/2}~ 1p_{ 1/2}$ &  1 &  0 & -0.742 \\
 $ 1p_{ 3/2}~ 1p_{ 1/2}~ 1p_{ 3/2}~ 0f_{ 5/2}$ &  1 &  0 & -0.428 \\
 $ 1p_{ 3/2}~ 0f_{ 5/2}~ 1p_{ 3/2}~ 0f_{ 5/2}$ &  1 &  0 & -1.046 \\
 $ 0f_{ 7/2}~ 1p_{ 3/2}~ 0f_{ 7/2}~ 1p_{ 3/2}$ &  2 &  0 & -0.951 \\
 $ 0f_{ 7/2}~ 1p_{ 3/2}~ 0f_{ 7/2}~ 0f_{ 5/2}$ &  2 &  0 & -0.738 \\
 $ 0f_{ 7/2}~ 1p_{ 3/2}~ 1p_{ 3/2}~ 1p_{ 3/2}$ &  2 &  0 & -0.432 \\
 $ 0f_{ 7/2}~ 1p_{ 3/2}~ 1p_{ 3/2}~ 1p_{ 1/2}$ &  2 &  0 & -0.758 \\
 $ 0f_{ 7/2}~ 1p_{ 3/2}~ 1p_{ 3/2}~ 0f_{ 5/2}$ &  2 &  0 & -0.877 \\
 $ 0f_{ 7/2}~ 0f_{ 5/2}~ 0f_{ 7/2}~ 0f_{ 5/2}$ &  2 &  0 & -1.424 \\
 $ 0f_{ 7/2}~ 0f_{ 5/2}~ 1p_{ 3/2}~ 1p_{ 3/2}$ &  2 &  0 & -0.109 \\
 $ 0f_{ 7/2}~ 0f_{ 5/2}~ 1p_{ 3/2}~ 1p_{ 1/2}$ &  2 &  0 & -0.641 \\
 $ 0f_{ 7/2}~ 0f_{ 5/2}~ 1p_{ 3/2}~ 0f_{ 5/2}$ &  2 &  0 & -0.627 \\
 $ 1p_{ 3/2}~ 1p_{ 3/2}~ 1p_{ 3/2}~ 1p_{ 3/2}$ &  2 &  0 & -0.547 \\
 $ 1p_{ 3/2}~ 1p_{ 3/2}~ 1p_{ 3/2}~ 1p_{ 1/2}$ &  2 &  0 & -0.688 \\
 $ 1p_{ 3/2}~ 1p_{ 3/2}~ 1p_{ 3/2}~ 0f_{ 5/2}$ &  2 &  0 & -0.165 \\
 $ 1p_{ 3/2}~ 1p_{ 1/2}~ 1p_{ 3/2}~ 1p_{ 1/2}$ &  2 &  0 & -1.082 \\
 $ 1p_{ 3/2}~ 1p_{ 1/2}~ 1p_{ 3/2}~ 0f_{ 5/2}$ &  2 &  0 & -0.443 \\
 $ 1p_{ 3/2}~ 0f_{ 5/2}~ 1p_{ 3/2}~ 0f_{ 5/2}$ &  2 &  0 & -0.447 \\
 $ 0f_{ 7/2}~ 1p_{ 3/2}~ 0f_{ 7/2}~ 1p_{ 3/2}$ &  3 &  0 & -0.348 \\
 $ 0f_{ 7/2}~ 1p_{ 3/2}~ 0f_{ 7/2}~ 1p_{ 1/2}$ &  3 &  0 &  0.587 \\
 $ 0f_{ 7/2}~ 1p_{ 3/2}~ 0f_{ 7/2}~ 0f_{ 5/2}$ &  3 &  0 &  0.141 \\
 $ 0f_{ 7/2}~ 1p_{ 3/2}~ 1p_{ 3/2}~ 1p_{ 3/2}$ &  3 &  0 & -0.449 \\
 $ 0f_{ 7/2}~ 1p_{ 3/2}~ 1p_{ 3/2}~ 0f_{ 5/2}$ &  3 &  0 &  0.141 \\
 $ 0f_{ 7/2}~ 1p_{ 1/2}~ 0f_{ 7/2}~ 1p_{ 1/2}$ &  3 &  0 & -0.618 \\
 $ 0f_{ 7/2}~ 1p_{ 1/2}~ 0f_{ 7/2}~ 0f_{ 5/2}$ &  3 &  0 & -0.275 \\
 $ 0f_{ 7/2}~ 1p_{ 1/2}~ 1p_{ 3/2}~ 1p_{ 3/2}$ &  3 &  0 &  0.528 \\
 $ 0f_{ 7/2}~ 1p_{ 1/2}~ 1p_{ 3/2}~ 0f_{ 5/2}$ &  3 &  0 & -0.117 \\
 $ 0f_{ 7/2}~ 0f_{ 5/2}~ 0f_{ 7/2}~ 0f_{ 5/2}$ &  3 &  0 & -0.524 \\
 $ 0f_{ 7/2}~ 0f_{ 5/2}~ 1p_{ 3/2}~ 1p_{ 3/2}$ &  3 &  0 &  0.436 \\
 $ 0f_{ 7/2}~ 0f_{ 5/2}~ 1p_{ 3/2}~ 0f_{ 5/2}$ &  3 &  0 & -0.369 \\
 $ 1p_{ 3/2}~ 1p_{ 3/2}~ 1p_{ 3/2}~ 1p_{ 3/2}$ &  3 &  0 & -0.811 \\
 $ 1p_{ 3/2}~ 1p_{ 3/2}~ 1p_{ 3/2}~ 0f_{ 5/2}$ &  3 &  0 &  0.227 \\
 $ 1p_{ 3/2}~ 0f_{ 5/2}~ 1p_{ 3/2}~ 0f_{ 5/2}$ &  3 &  0 & -0.261 \\
 $ 0f_{ 7/2}~ 1p_{ 3/2}~ 0f_{ 7/2}~ 1p_{ 3/2}$ &  4 &  0 & -0.100 \\
 $ 0f_{ 7/2}~ 1p_{ 3/2}~ 0f_{ 7/2}~ 1p_{ 1/2}$ &  4 &  0 & -0.297 \\
 $ 0f_{ 7/2}~ 1p_{ 3/2}~ 0f_{ 7/2}~ 0f_{ 5/2}$ &  4 &  0 & -0.229 \\
 $ 0f_{ 7/2}~ 1p_{ 3/2}~ 1p_{ 3/2}~ 0f_{ 5/2}$ &  4 &  0 & -0.615 \\
 $ 0f_{ 7/2}~ 1p_{ 1/2}~ 0f_{ 7/2}~ 1p_{ 1/2}$ &  4 &  0 & -0.592 \\
 $ 0f_{ 7/2}~ 1p_{ 1/2}~ 0f_{ 7/2}~ 0f_{ 5/2}$ &  4 &  0 & -0.590 \\
 $ 0f_{ 7/2}~ 1p_{ 1/2}~ 1p_{ 3/2}~ 0f_{ 5/2}$ &  4 &  0 & -1.004 \\
 $ 0f_{ 7/2}~ 0f_{ 5/2}~ 0f_{ 7/2}~ 0f_{ 5/2}$ &  4 &  0 & -0.848 \\
 $ 0f_{ 7/2}~ 0f_{ 5/2}~ 1p_{ 3/2}~ 0f_{ 5/2}$ &  4 &  0 & -0.783 \\
 $ 1p_{ 3/2}~ 0f_{ 5/2}~ 1p_{ 3/2}~ 0f_{ 5/2}$ &  4 &  0 & -0.753 \\
 $ 0f_{ 7/2}~ 1p_{ 3/2}~ 0f_{ 7/2}~ 1p_{ 3/2}$ &  5 &  0 & -0.884 \\
 $ 0f_{ 7/2}~ 1p_{ 3/2}~ 0f_{ 7/2}~ 0f_{ 5/2}$ &  5 &  0 &  0.274 \\
 $ 0f_{ 7/2}~ 0f_{ 5/2}~ 0f_{ 7/2}~ 0f_{ 5/2}$ &  5 &  0 & -0.179 \\
 $ 0f_{ 7/2}~ 0f_{ 5/2}~ 0f_{ 7/2}~ 0f_{ 5/2}$ &  6 &  0 & -1.577 \\
 $ 0f_{ 7/2}~ 0g_{ 9/2}~ 0f_{ 7/2}~ 0g_{ 9/2}$ &  1 &  0 & -0.846 \\
 $ 0f_{ 7/2}~ 0g_{ 9/2}~ 0f_{ 7/2}~ 0g_{ 9/2}$ &  2 &  0 & -0.617 \\
 $ 0f_{ 7/2}~ 0g_{ 9/2}~ 0f_{ 7/2}~ 0g_{ 9/2}$ &  3 &  0 & -0.176 \\
 $ 0f_{ 7/2}~ 0g_{ 9/2}~ 1p_{ 3/2}~ 0g_{ 9/2}$ &  3 &  0 & -0.419 \\
 $ 1p_{ 3/2}~ 0g_{ 9/2}~ 1p_{ 3/2}~ 0g_{ 9/2}$ &  3 &  0 & -0.723 \\
 $ 0f_{ 7/2}~ 0g_{ 9/2}~ 0f_{ 7/2}~ 0g_{ 9/2}$ &  4 &  0 & -0.091 \\
 $ 0f_{ 7/2}~ 0g_{ 9/2}~ 1p_{ 3/2}~ 0g_{ 9/2}$ &  4 &  0 & -0.419 \\
 $ 1p_{ 3/2}~ 0g_{ 9/2}~ 1p_{ 3/2}~ 0g_{ 9/2}$ &  4 &  0 & -0.211 \\
 $ 0f_{ 7/2}~ 0g_{ 9/2}~ 0f_{ 7/2}~ 0g_{ 9/2}$ &  5 &  0 & -0.003 \\
 $ 0f_{ 7/2}~ 0g_{ 9/2}~ 1p_{ 3/2}~ 0g_{ 9/2}$ &  5 &  0 & -0.282 \\
$ 1p_{ 3/2}~ 0g_{ 9/2}~ 1p_{ 3/2}~ 0g_{ 9/2}$ &  5 &  0 & -0.043 \\
 $ 0f_{ 7/2}~ 0g_{ 9/2}~ 0f_{ 7/2}~ 0g_{ 9/2}$ &  6 &  0 & -0.149 \\
 $ 0f_{ 7/2}~ 0g_{ 9/2}~ 1p_{ 3/2}~ 0g_{ 9/2}$ &  6 &  0 & -0.574 \\
 $ 1p_{ 3/2}~ 0g_{ 9/2}~ 1p_{ 3/2}~ 0g_{ 9/2}$ &  6 &  0 & -0.533 \\
 $ 0f_{ 7/2}~ 0g_{ 9/2}~ 0f_{ 7/2}~ 0g_{ 9/2}$ &  7 &  0 &  0.073 \\
 $ 0f_{ 7/2}~ 0g_{ 9/2}~ 0f_{ 7/2}~ 0g_{ 9/2}$ &  8 &  0 & -1.134 \\
\end{tabular}
\end{ruledtabular}
\end{table}

\begin{table}[H]
\caption{Proton-proton, neutron-neutron, and proton-neutron matrix
  elements (in MeV) derived for calculations in model space (II). They
  are antisymmetrized, and normalized by a factor $1/ \sqrt{ (1 +
    \delta_{j_aj_b})(1 + \delta_{j_cj_d})}$.}
\begin{ruledtabular}
\begin{tabular}{cccc}
$n_a l_a j_a ~ n_b l_b j_b ~ n_c l_c j_c ~ n_d l_d j_d $ & $J$ & $T_z$
  &  TBME \\
\colrule
 $ 0f_{ 7/2}~ 0f_{ 7/2}~ 0f_{ 7/2}~ 0f_{ 7/2}$ &  0 &  1 & -2.026 \\
 $ 0f_{ 7/2}~ 0f_{ 7/2}~ 1p_{ 3/2}~ 1p_{ 3/2}$ &  0 &  1 & -1.225 \\
 $ 1p_{ 3/2}~ 1p_{ 3/2}~ 1p_{ 3/2}~ 1p_{ 3/2}$ &  0 &  1 & -0.068 \\
 $ 0f_{ 7/2}~ 0f_{ 7/2}~ 0f_{ 7/2}~ 0f_{ 7/2}$ &  2 &  1 & -0.416 \\
 $ 0f_{ 7/2}~ 0f_{ 7/2}~ 0f_{ 7/2}~ 1p_{ 3/2}$ &  2 &  1 & -0.710 \\
 $ 0f_{ 7/2}~ 0f_{ 7/2}~ 1p_{ 3/2}~ 1p_{ 3/2}$ &  2 &  1 & -0.358 \\
 $ 0f_{ 7/2}~ 1p_{ 3/2}~ 0f_{ 7/2}~ 1p_{ 3/2}$ &  2 &  1 & -0.556 \\
 $ 0f_{ 7/2}~ 1p_{ 3/2}~ 1p_{ 3/2}~ 1p_{ 3/2}$ &  2 &  1 & -0.361 \\
 $ 1p_{ 3/2}~ 1p_{ 3/2}~ 1p_{ 3/2}~ 1p_{ 3/2}$ &  2 &  1 &  0.104 \\
 $ 0f_{ 7/2}~ 1p_{ 3/2}~ 0f_{ 7/2}~ 1p_{ 3/2}$ &  3 &  1 &  0.329 \\
 $ 0f_{ 7/2}~ 0f_{ 7/2}~ 0f_{ 7/2}~ 0f_{ 7/2}$ &  4 &  1 &  0.133 \\
 $ 0f_{ 7/2}~ 0f_{ 7/2}~ 0f_{ 7/2}~ 1p_{ 3/2}$ &  4 &  1 & -0.410 \\
 $ 0f_{ 7/2}~ 1p_{ 3/2}~ 0f_{ 7/2}~ 1p_{ 3/2}$ &  4 &  1 &  0.057 \\
 $ 0f_{ 7/2}~ 1p_{ 3/2}~ 0f_{ 7/2}~ 1p_{ 3/2}$ &  5 &  1 &  0.496 \\
 $ 0f_{ 7/2}~ 0f_{ 7/2}~ 0f_{ 7/2}~ 0f_{ 7/2}$ &  6 &  1 &  0.407 \\
 $ 1p_{ 3/2}~ 1p_{ 3/2}~ 1p_{ 3/2}~ 1p_{ 3/2}$ &  0 & -1 & -0.715 \\
 $ 1p_{ 3/2}~ 1p_{ 3/2}~ 1p_{ 1/2}~ 1p_{ 1/2}$ &  0 & -1 & -0.869 \\
 $ 1p_{ 3/2}~ 1p_{ 3/2}~ 0f_{ 5/2}~ 0f_{ 5/2}$ &  0 & -1 & -0.695 \\
 $ 1p_{ 3/2}~ 1p_{ 3/2}~ 0g_{ 9/2}~ 0g_{ 9/2}$ &  0 & -1 &  1.026 \\
 $ 1p_{ 3/2}~ 1p_{ 3/2}~ 1d_{ 5/2}~ 1d_{ 5/2}$ &  0 & -1 &  0.967 \\
 $ 1p_{ 1/2}~ 1p_{ 1/2}~ 1p_{ 1/2}~ 1p_{ 1/2}$ &  0 & -1 & -0.302 \\
 $ 1p_{ 1/2}~ 1p_{ 1/2}~ 0f_{ 5/2}~ 0f_{ 5/2}$ &  0 & -1 & -0.470 \\
 $ 1p_{ 1/2}~ 1p_{ 1/2}~ 0g_{ 9/2}~ 0g_{ 9/2}$ &  0 & -1 &  0.846 \\
 $ 1p_{ 1/2}~ 1p_{ 1/2}~ 1d_{ 5/2}~ 1d_{ 5/2}$ &  0 & -1 &  0.603 \\
 $ 0f_{ 5/2}~ 0f_{ 5/2}~ 0f_{ 5/2}~ 0f_{ 5/2}$ &  0 & -1 & -0.978 \\
 $ 0f_{ 5/2}~ 0f_{ 5/2}~ 0g_{ 9/2}~ 0g_{ 9/2}$ &  0 & -1 &  1.555 \\
 $ 0f_{ 5/2}~ 0f_{ 5/2}~ 1d_{ 5/2}~ 1d_{ 5/2}$ &  0 & -1 &  0.610 \\
 $ 0g_{ 9/2}~ 0g_{ 9/2}~ 0g_{ 9/2}~ 0g_{ 9/2}$ &  0 & -1 & -1.171 \\
 $ 0g_{ 9/2}~ 0g_{ 9/2}~ 1d_{ 5/2}~ 1d_{ 5/2}$ &  0 & -1 & -0.884 \\
 $ 1d_{ 5/2}~ 1d_{ 5/2}~ 1d_{ 5/2}~ 1d_{ 5/2}$ &  0 & -1 & -0.904 \\
 $ 1p_{ 3/2}~ 1p_{ 1/2}~ 1p_{ 3/2}~ 1p_{ 1/2}$ &  1 & -1 &  0.165 \\
 $ 1p_{ 3/2}~ 1p_{ 1/2}~ 1p_{ 3/2}~ 0f_{ 5/2}$ &  1 & -1 & -0.053 \\
 $ 1p_{ 3/2}~ 0f_{ 5/2}~ 1p_{ 3/2}~ 0f_{ 5/2}$ &  1 & -1 &  0.050 \\
 $ 1p_{ 3/2}~ 1p_{ 3/2}~ 1p_{ 3/2}~ 1p_{ 3/2}$ &  2 & -1 & -0.219 \\
 $ 1p_{ 3/2}~ 1p_{ 3/2}~ 1p_{ 3/2}~ 1p_{ 1/2}$ &  2 & -1 & -0.372 \\
 $ 1p_{ 3/2}~ 1p_{ 3/2}~ 1p_{ 3/2}~ 0f_{ 5/2}$ &  2 & -1 & -0.158 \\
 $ 1p_{ 3/2}~ 1p_{ 3/2}~ 1p_{ 1/2}~ 0f_{ 5/2}$ &  2 & -1 & -0.163 \\
 $ 1p_{ 3/2}~ 1p_{ 3/2}~ 0f_{ 5/2}~ 0f_{ 5/2}$ &  2 & -1 & -0.168 \\
 $ 1p_{ 3/2}~ 1p_{ 3/2}~ 0g_{ 9/2}~ 0g_{ 9/2}$ &  2 & -1 &  0.393 \\
 $ 1p_{ 3/2}~ 1p_{ 3/2}~ 0g_{ 9/2}~ 1d_{ 5/2}$ &  2 & -1 &  0.340 \\
 $ 1p_{ 3/2}~ 1p_{ 3/2}~ 1d_{ 5/2}~ 1d_{ 5/2}$ &  2 & -1 &  0.471 \\
 $ 1p_{ 3/2}~ 1p_{ 1/2}~ 1p_{ 3/2}~ 1p_{ 1/2}$ &  2 & -1 & -0.418 \\
 $ 1p_{ 3/2}~ 1p_{ 1/2}~ 1p_{ 3/2}~ 0f_{ 5/2}$ &  2 & -1 & -0.172 \\
 $ 1p_{ 3/2}~ 1p_{ 1/2}~ 1p_{ 1/2}~ 0f_{ 5/2}$ &  2 & -1 & -0.384 \\
 $ 1p_{ 3/2}~ 1p_{ 1/2}~ 0f_{ 5/2}~ 0f_{ 5/2}$ &  2 & -1 & -0.280 \\
 $ 1p_{ 3/2}~ 1p_{ 1/2}~ 0g_{ 9/2}~ 0g_{ 9/2}$ &  2 & -1 &  0.271 \\
 $ 1p_{ 3/2}~ 1p_{ 1/2}~ 0g_{ 9/2}~ 1d_{ 5/2}$ &  2 & -1 &  0.536 \\
 $ 1p_{ 3/2}~ 1p_{ 1/2}~ 1d_{ 5/2}~ 1d_{ 5/2}$ &  2 & -1 &  0.207 \\
 $ 1p_{ 3/2}~ 0f_{ 5/2}~ 1p_{ 3/2}~ 0f_{ 5/2}$ &  2 & -1 &  0.072 \\
 $ 1p_{ 3/2}~ 0f_{ 5/2}~ 1p_{ 1/2}~ 0f_{ 5/2}$ &  2 & -1 & -0.277 \\
 $ 1p_{ 3/2}~ 0f_{ 5/2}~ 0f_{ 5/2}~ 0f_{ 5/2}$ &  2 & -1 & -0.101 \\
 $ 1p_{ 3/2}~ 0f_{ 5/2}~ 0g_{ 9/2}~ 0g_{ 9/2}$ &  2 & -1 &  0.475 \\
 $ 1p_{ 3/2}~ 0f_{ 5/2}~ 0g_{ 9/2}~ 1d_{ 5/2}$ &  2 & -1 &  0.072 \\
 $ 1p_{ 3/2}~ 0f_{ 5/2}~ 1d_{ 5/2}~ 1d_{ 5/2}$ &  2 & -1 &  0.089 \\
 $ 1p_{ 1/2}~ 0f_{ 5/2}~ 1p_{ 1/2}~ 0f_{ 5/2}$ &  2 & -1 & -0.277 \\
 $ 1p_{ 1/2}~ 0f_{ 5/2}~ 0f_{ 5/2}~ 0f_{ 5/2}$ &  2 & -1 & -0.273 \\
 $ 1p_{ 1/2}~ 0f_{ 5/2}~ 0g_{ 9/2}~ 0g_{ 9/2}$ &  2 & -1 &  0.552 \\
\end{tabular}
\end{ruledtabular}
\label{tbmeII}
\end{table}

\begin{table}[H]
\begin{ruledtabular}
\begin{tabular}{cccc}
\colrule
 $ 1p_{ 1/2}~ 0f_{ 5/2}~ 0g_{ 9/2}~ 1d_{ 5/2}$ &  2 & -1 &  0.526 \\
 $ 1p_{ 1/2}~ 0f_{ 5/2}~ 1d_{ 5/2}~ 1d_{ 5/2}$ &  2 & -1 &  0.186 \\
 $ 0f_{ 5/2}~ 0f_{ 5/2}~ 0f_{ 5/2}~ 0f_{ 5/2}$ &  2 & -1 & -0.400 \\
 $ 0f_{ 5/2}~ 0f_{ 5/2}~ 0g_{ 9/2}~ 0g_{ 9/2}$ &  2 & -1 &  0.229 \\
 $ 0f_{ 5/2}~ 0f_{ 5/2}~ 0g_{ 9/2}~ 1d_{ 5/2}$ &  2 & -1 &  0.269 \\
 $ 0f_{ 5/2}~ 0f_{ 5/2}~ 1d_{ 5/2}~ 1d_{ 5/2}$ &  2 & -1 &  0.138 \\
 $ 0g_{ 9/2}~ 0g_{ 9/2}~ 0g_{ 9/2}~ 0g_{ 9/2}$ &  2 & -1 & -0.756 \\
 $ 0g_{ 9/2}~ 0g_{ 9/2}~ 0g_{ 9/2}~ 1d_{ 5/2}$ &  2 & -1 & -0.377 \\
 $ 0g_{ 9/2}~ 0g_{ 9/2}~ 1d_{ 5/2}~ 1d_{ 5/2}$ &  2 & -1 & -0.347 \\
 $ 0g_{ 9/2}~ 1d_{ 5/2}~ 0g_{ 9/2}~ 1d_{ 5/2}$ &  2 & -1 & -0.774 \\
 $ 0g_{ 9/2}~ 1d_{ 5/2}~ 1d_{ 5/2}~ 1d_{ 5/2}$ &  2 & -1 & -0.361 \\
 $ 1d_{ 5/2}~ 1d_{ 5/2}~ 1d_{ 5/2}~ 1d_{ 5/2}$ &  2 & -1 & -0.275 \\
 $ 1p_{ 3/2}~ 0f_{ 5/2}~ 1p_{ 3/2}~ 0f_{ 5/2}$ &  3 & -1 &  0.126 \\
 $ 1p_{ 3/2}~ 0f_{ 5/2}~ 1p_{ 1/2}~ 0f_{ 5/2}$ &  3 & -1 & -0.005 \\
 $ 1p_{ 3/2}~ 0f_{ 5/2}~ 0g_{ 9/2}~ 1d_{ 5/2}$ &  3 & -1 &  0.069 \\
 $ 1p_{ 1/2}~ 0f_{ 5/2}~ 1p_{ 1/2}~ 0f_{ 5/2}$ &  3 & -1 &  0.172 \\
 $ 1p_{ 1/2}~ 0f_{ 5/2}~ 0g_{ 9/2}~ 1d_{ 5/2}$ &  3 & -1 &  0.125 \\
 $ 0g_{ 9/2}~ 1d_{ 5/2}~ 0g_{ 9/2}~ 1d_{ 5/2}$ &  3 & -1 & -0.138 \\
 $ 1p_{ 3/2}~ 0f_{ 5/2}~ 1p_{ 3/2}~ 0f_{ 5/2}$ &  4 & -1 & -0.326 \\
 $ 1p_{ 3/2}~ 0f_{ 5/2}~ 0f_{ 5/2}~ 0f_{ 5/2}$ &  4 & -1 & -0.244 \\
 $ 1p_{ 3/2}~ 0f_{ 5/2}~ 0g_{ 9/2}~ 0g_{ 9/2}$ &  4 & -1 &  0.469 \\
 $ 1p_{ 3/2}~ 0f_{ 5/2}~ 0g_{ 9/2}~ 1d_{ 5/2}$ &  4 & -1 &  0.350 \\
 $ 1p_{ 3/2}~ 0f_{ 5/2}~ 1d_{ 5/2}~ 1d_{ 5/2}$ &  4 & -1 &  0.089 \\
 $ 0f_{ 5/2}~ 0f_{ 5/2}~ 0f_{ 5/2}~ 0f_{ 5/2}$ &  4 & -1 & -0.086 \\
 $ 0f_{ 5/2}~ 0f_{ 5/2}~ 0g_{ 9/2}~ 0g_{ 9/2}$ &  4 & -1 &  0.109 \\
 $ 0f_{ 5/2}~ 0f_{ 5/2}~ 0g_{ 9/2}~ 1d_{ 5/2}$ &  4 & -1 &  0.097 \\
 $ 0f_{ 5/2}~ 0f_{ 5/2}~ 1d_{ 5/2}~ 1d_{ 5/2}$ &  4 & -1 &  0.089 \\
 $ 0g_{ 9/2}~ 0g_{ 9/2}~ 0g_{ 9/2}~ 0g_{ 9/2}$ &  4 & -1 & -0.273 \\
 $ 0g_{ 9/2}~ 0g_{ 9/2}~ 0g_{ 9/2}~ 1d_{ 5/2}$ &  4 & -1 & -0.357 \\
 $ 0g_{ 9/2}~ 0g_{ 9/2}~ 1d_{ 5/2}~ 1d_{ 5/2}$ &  4 & -1 & -0.156 \\
 $ 0g_{ 9/2}~ 1d_{ 5/2}~ 0g_{ 9/2}~ 1d_{ 5/2}$ &  4 & -1 & -0.229 \\
 $ 0g_{ 9/2}~ 1d_{ 5/2}~ 1d_{ 5/2}~ 1d_{ 5/2}$ &  4 & -1 & -0.204 \\
 $ 1d_{ 5/2}~ 1d_{ 5/2}~ 1d_{ 5/2}~ 1d_{ 5/2}$ &  4 & -1 & -0.072 \\
 $ 0g_{ 9/2}~ 1d_{ 5/2}~ 0g_{ 9/2}~ 1d_{ 5/2}$ &  5 & -1 &  0.121 \\
 $ 0g_{ 9/2}~ 0g_{ 9/2}~ 0g_{ 9/2}~ 0g_{ 9/2}$ &  6 & -1 & -0.105 \\
 $ 0g_{ 9/2}~ 0g_{ 9/2}~ 0g_{ 9/2}~ 1d_{ 5/2}$ &  6 & -1 & -0.234 \\
 $ 0g_{ 9/2}~ 1d_{ 5/2}~ 0g_{ 9/2}~ 1d_{ 5/2}$ &  6 & -1 & -0.031 \\
 $ 0g_{ 9/2}~ 1d_{ 5/2}~ 0g_{ 9/2}~ 1d_{ 5/2}$ &  7 & -1 &  0.127 \\
 $ 0g_{ 9/2}~ 0g_{ 9/2}~ 0g_{ 9/2}~ 0g_{ 9/2}$ &  8 & -1 & -0.001 \\
 $ 0f_{ 5/2}~ 1d_{ 5/2}~ 0f_{ 5/2}~ 1d_{ 5/2}$ &  0 & -1 & -0.497 \\
 $ 1p_{ 3/2}~ 1d_{ 5/2}~ 1p_{ 3/2}~ 1d_{ 5/2}$ &  1 & -1 & -0.765 \\
 $ 1p_{ 3/2}~ 1d_{ 5/2}~ 0f_{ 5/2}~ 1d_{ 5/2}$ &  1 & -1 &  0.246 \\
 $ 0f_{ 5/2}~ 1d_{ 5/2}~ 0f_{ 5/2}~ 1d_{ 5/2}$ &  1 & -1 & -0.089 \\
 $ 1p_{ 3/2}~ 1d_{ 5/2}~ 1p_{ 3/2}~ 1d_{ 5/2}$ &  2 & -1 & -0.080 \\
 $ 1p_{ 3/2}~ 1d_{ 5/2}~ 1p_{ 1/2}~ 1d_{ 5/2}$ &  2 & -1 & -0.103 \\
 $ 1p_{ 3/2}~ 1d_{ 5/2}~ 0f_{ 5/2}~ 0g_{ 9/2}$ &  2 & -1 & -0.293 \\
 $ 1p_{ 3/2}~ 1d_{ 5/2}~ 0f_{ 5/2}~ 1d_{ 5/2}$ &  2 & -1 &  0.032 \\
 $ 1p_{ 1/2}~ 1d_{ 5/2}~ 1p_{ 1/2}~ 1d_{ 5/2}$ &  2 & -1 & -0.072 \\
 $ 1p_{ 1/2}~ 1d_{ 5/2}~ 0f_{ 5/2}~ 0g_{ 9/2}$ &  2 & -1 & -0.348 \\
 $ 1p_{ 1/2}~ 1d_{ 5/2}~ 0f_{ 5/2}~ 1d_{ 5/2}$ &  2 & -1 & -0.123 \\
 $ 0f_{ 5/2}~ 0g_{ 9/2}~ 0f_{ 5/2}~ 0g_{ 9/2}$ &  2 & -1 & -0.382 \\
 $ 0f_{ 5/2}~ 0g_{ 9/2}~ 0f_{ 5/2}~ 1d_{ 5/2}$ &  2 & -1 & -0.216 \\
 $ 0f_{ 5/2}~ 1d_{ 5/2}~ 0f_{ 5/2}~ 1d_{ 5/2}$ &  2 & -1 & -0.057 \\
 $ 1p_{ 3/2}~ 0g_{ 9/2}~ 1p_{ 3/2}~ 0g_{ 9/2}$ &  3 & -1 & -0.497 \\
 $ 1p_{ 3/2}~ 0g_{ 9/2}~ 1p_{ 3/2}~ 1d_{ 5/2}$ &  3 & -1 & -0.337 \\
 $ 1p_{ 3/2}~ 0g_{ 9/2}~ 1p_{ 1/2}~ 1d_{ 5/2}$ &  3 & -1 &  0.420 \\
 $ 1p_{ 3/2}~ 0g_{ 9/2}~ 0f_{ 5/2}~ 0g_{ 9/2}$ &  3 & -1 &  0.199 \\
 $ 1p_{ 3/2}~ 0g_{ 9/2}~ 0f_{ 5/2}~ 1d_{ 5/2}$ &  3 & -1 &  0.285 \\
 $ 1p_{ 3/2}~ 1d_{ 5/2}~ 1p_{ 3/2}~ 1d_{ 5/2}$ &  3 & -1 & -0.386 \\
 $ 1p_{ 3/2}~ 1d_{ 5/2}~ 1p_{ 1/2}~ 1d_{ 5/2}$ &  3 & -1 &  0.315 \\
 $ 1p_{ 3/2}~ 1d_{ 5/2}~ 0f_{ 5/2}~ 0g_{ 9/2}$ &  3 & -1 &  0.023 \\
 $ 1p_{ 3/2}~ 1d_{ 5/2}~ 0f_{ 5/2}~ 1d_{ 5/2}$ &  3 & -1 &  0.182 \\
 $ 1p_{ 1/2}~ 1d_{ 5/2}~ 1p_{ 1/2}~ 1d_{ 5/2}$ &  3 & -1 & -0.416 \\
 $ 1p_{ 1/2}~ 1d_{ 5/2}~ 0f_{ 5/2}~ 0g_{ 9/2}$ &  3 & -1 & -0.274 \\
 $ 1p_{ 1/2}~ 1d_{ 5/2}~ 0f_{ 5/2}~ 1d_{ 5/2}$ &  3 & -1 & -0.240 \\
 $ 0f_{ 5/2}~ 0g_{ 9/2}~ 0f_{ 5/2}~ 0g_{ 9/2}$ &  3 & -1 & -0.194 \\
 $ 0f_{ 5/2}~ 0g_{ 9/2}~ 0f_{ 5/2}~ 1d_{ 5/2}$ &  3 & -1 & -0.300 \\
 $ 0f_{ 5/2}~ 1d_{ 5/2}~ 0f_{ 5/2}~ 1d_{ 5/2}$ &  3 & -1 & -0.014 \\
\end{tabular}
\end{ruledtabular}
\end{table}

\begin{table}[H]
\begin{ruledtabular}
\begin{tabular}{cccc}
\colrule
 $ 1p_{ 3/2}~ 0g_{ 9/2}~ 1p_{ 3/2}~ 0g_{ 9/2}$ &  4 & -1 &  0.054 \\
 $ 1p_{ 3/2}~ 0g_{ 9/2}~ 1p_{ 3/2}~ 1d_{ 5/2}$ &  4 & -1 & -0.024 \\
 $ 1p_{ 3/2}~ 0g_{ 9/2}~ 1p_{ 1/2}~ 0g_{ 9/2}$ &  4 & -1 & -0.105 \\
 $ 1p_{ 3/2}~ 0g_{ 9/2}~ 0f_{ 5/2}~ 0g_{ 9/2}$ &  4 & -1 &  0.151 \\
 $ 1p_{ 3/2}~ 0g_{ 9/2}~ 0f_{ 5/2}~ 1d_{ 5/2}$ &  4 & -1 &  0.076 \\
 $ 1p_{ 3/2}~ 1d_{ 5/2}~ 1p_{ 3/2}~ 1d_{ 5/2}$ &  4 & -1 &  0.257 \\
 $ 1p_{ 3/2}~ 1d_{ 5/2}~ 1p_{ 1/2}~ 0g_{ 9/2}$ &  4 & -1 & -0.035 \\
 $ 1p_{ 3/2}~ 1d_{ 5/2}~ 0f_{ 5/2}~ 0g_{ 9/2}$ &  4 & -1 & -0.128 \\
 $ 1p_{ 3/2}~ 1d_{ 5/2}~ 0f_{ 5/2}~ 1d_{ 5/2}$ &  4 & -1 & -0.018 \\
 $ 1p_{ 1/2}~ 0g_{ 9/2}~ 1p_{ 1/2}~ 0g_{ 9/2}$ &  4 & -1 &  0.113 \\
 $ 1p_{ 1/2}~ 0g_{ 9/2}~ 0f_{ 5/2}~ 0g_{ 9/2}$ &  4 & -1 & -0.009 \\
 $ 1p_{ 1/2}~ 0g_{ 9/2}~ 0f_{ 5/2}~ 1d_{ 5/2}$ &  4 & -1 &  0.022 \\
 $ 0f_{ 5/2}~ 0g_{ 9/2}~ 0f_{ 5/2}~ 0g_{ 9/2}$ &  4 & -1 &  0.116 \\
 $ 0f_{ 5/2}~ 0g_{ 9/2}~ 0f_{ 5/2}~ 1d_{ 5/2}$ &  4 & -1 & -0.137 \\
 $ 0f_{ 5/2}~ 1d_{ 5/2}~ 0f_{ 5/2}~ 1d_{ 5/2}$ &  4 & -1 &  0.027 \\
 $ 1p_{ 3/2}~ 0g_{ 9/2}~ 1p_{ 3/2}~ 0g_{ 9/2}$ &  5 & -1 & -0.075 \\
 $ 1p_{ 3/2}~ 0g_{ 9/2}~ 1p_{ 1/2}~ 0g_{ 9/2}$ &  5 & -1 &  0.333 \\
 $ 1p_{ 3/2}~ 0g_{ 9/2}~ 0f_{ 5/2}~ 0g_{ 9/2}$ &  5 & -1 &  0.136 \\
 $ 1p_{ 3/2}~ 0g_{ 9/2}~ 0f_{ 5/2}~ 1d_{ 5/2}$ &  5 & -1 &  0.347 \\
 $ 1p_{ 1/2}~ 0g_{ 9/2}~ 1p_{ 1/2}~ 0g_{ 9/2}$ &  5 & -1 & -0.309 \\
 $ 1p_{ 1/2}~ 0g_{ 9/2}~ 0f_{ 5/2}~ 0g_{ 9/2}$ &  5 & -1 & -0.233 \\
 $ 1p_{ 1/2}~ 0g_{ 9/2}~ 0f_{ 5/2}~ 1d_{ 5/2}$ &  5 & -1 & -0.449 \\
 $ 0f_{ 5/2}~ 0g_{ 9/2}~ 0f_{ 5/2}~ 0g_{ 9/2}$ &  5 & -1 & -0.050 \\
 $ 0f_{ 5/2}~ 0g_{ 9/2}~ 0f_{ 5/2}~ 1d_{ 5/2}$ &  5 & -1 & -0.418 \\
 $ 0f_{ 5/2}~ 1d_{ 5/2}~ 0f_{ 5/2}~ 1d_{ 5/2}$ &  5 & -1 & -0.373 \\
 $ 1p_{ 3/2}~ 0g_{ 9/2}~ 1p_{ 3/2}~ 0g_{ 9/2}$ &  6 & -1 &  0.267 \\
 $ 1p_{ 3/2}~ 0g_{ 9/2}~ 0f_{ 5/2}~ 0g_{ 9/2}$ &  6 & -1 &  0.176 \\
 $ 0f_{ 5/2}~ 0g_{ 9/2}~ 0f_{ 5/2}~ 0g_{ 9/2}$ &  6 & -1 &  0.158 \\
 $ 0f_{ 5/2}~ 0g_{ 9/2}~ 0f_{ 5/2}~ 0g_{ 9/2}$ &  7 & -1 & -0.719 \\
 $ 1p_{ 3/2}~ 1p_{ 3/2}~ 1p_{ 3/2}~ 1p_{ 3/2}$ &  0 &  0 & -2.101 \\
 $ 0f_{ 7/2}~ 0f_{ 5/2}~ 0f_{ 7/2}~ 0f_{ 5/2}$ &  1 &  0 & -2.000 \\
 $ 0f_{ 7/2}~ 0f_{ 5/2}~ 1p_{ 3/2}~ 1p_{ 3/2}$ &  1 &  0 &  0.804 \\
 $ 0f_{ 7/2}~ 0f_{ 5/2}~ 1p_{ 3/2}~ 1p_{ 1/2}$ &  1 &  0 & -0.977 \\
 $ 0f_{ 7/2}~ 0f_{ 5/2}~ 1p_{ 3/2}~ 0f_{ 5/2}$ &  1 &  0 & -0.680 \\
 $ 1p_{ 3/2}~ 1p_{ 3/2}~ 1p_{ 3/2}~ 1p_{ 3/2}$ &  1 &  0 & -0.387 \\
 $ 1p_{ 3/2}~ 1p_{ 3/2}~ 1p_{ 3/2}~ 1p_{ 1/2}$ &  1 &  0 &  0.618 \\
 $ 1p_{ 3/2}~ 1p_{ 3/2}~ 1p_{ 3/2}~ 0f_{ 5/2}$ &  1 &  0 & -0.019 \\
 $ 1p_{ 3/2}~ 1p_{ 1/2}~ 1p_{ 3/2}~ 1p_{ 1/2}$ &  1 &  0 & -0.742 \\
 $ 1p_{ 3/2}~ 1p_{ 1/2}~ 1p_{ 3/2}~ 0f_{ 5/2}$ &  1 &  0 & -0.422 \\
 $ 1p_{ 3/2}~ 0f_{ 5/2}~ 1p_{ 3/2}~ 0f_{ 5/2}$ &  1 &  0 & -1.036 \\
 $ 0f_{ 7/2}~ 1p_{ 3/2}~ 0f_{ 7/2}~ 1p_{ 3/2}$ &  2 &  0 & -1.004 \\
 $ 0f_{ 7/2}~ 1p_{ 3/2}~ 0f_{ 7/2}~ 0f_{ 5/2}$ &  2 &  0 & -0.748 \\
 $ 0f_{ 7/2}~ 1p_{ 3/2}~ 1p_{ 3/2}~ 1p_{ 3/2}$ &  2 &  0 & -0.455 \\
 $ 0f_{ 7/2}~ 1p_{ 3/2}~ 1p_{ 3/2}~ 1p_{ 1/2}$ &  2 &  0 & -0.801 \\
 $ 0f_{ 7/2}~ 1p_{ 3/2}~ 1p_{ 3/2}~ 0f_{ 5/2}$ &  2 &  0 & -0.900 \\
 $ 0f_{ 7/2}~ 0f_{ 5/2}~ 0f_{ 7/2}~ 0f_{ 5/2}$ &  2 &  0 & -1.422 \\
 $ 0f_{ 7/2}~ 0f_{ 5/2}~ 1p_{ 3/2}~ 1p_{ 3/2}$ &  2 &  0 & -0.111 \\
 $ 0f_{ 7/2}~ 0f_{ 5/2}~ 1p_{ 3/2}~ 1p_{ 1/2}$ &  2 &  0 & -0.648 \\
 $ 0f_{ 7/2}~ 0f_{ 5/2}~ 1p_{ 3/2}~ 0f_{ 5/2}$ &  2 &  0 & -0.631 \\
 $ 1p_{ 3/2}~ 1p_{ 3/2}~ 1p_{ 3/2}~ 1p_{ 3/2}$ &  2 &  0 & -0.555 \\
 $ 1p_{ 3/2}~ 1p_{ 3/2}~ 1p_{ 3/2}~ 1p_{ 1/2}$ &  2 &  0 & -0.699 \\
 $ 1p_{ 3/2}~ 1p_{ 3/2}~ 1p_{ 3/2}~ 0f_{ 5/2}$ &  2 &  0 & -0.176 \\
 $ 1p_{ 3/2}~ 1p_{ 1/2}~ 1p_{ 3/2}~ 1p_{ 1/2}$ &  2 &  0 & -1.104 \\
 $ 1p_{ 3/2}~ 1p_{ 1/2}~ 1p_{ 3/2}~ 0f_{ 5/2}$ &  2 &  0 & -0.462 \\
 $ 1p_{ 3/2}~ 0f_{ 5/2}~ 1p_{ 3/2}~ 0f_{ 5/2}$ &  2 &  0 & -0.457 \\
 $ 0f_{ 7/2}~ 1p_{ 3/2}~ 0f_{ 7/2}~ 1p_{ 3/2}$ &  3 &  0 & -0.370 \\
 $ 0f_{ 7/2}~ 1p_{ 3/2}~ 0f_{ 7/2}~ 1p_{ 1/2}$ &  3 &  0 &  0.603 \\
 $ 0f_{ 7/2}~ 1p_{ 3/2}~ 0f_{ 7/2}~ 0f_{ 5/2}$ &  3 &  0 &  0.144 \\
 $ 0f_{ 7/2}~ 1p_{ 3/2}~ 1p_{ 3/2}~ 1p_{ 3/2}$ &  3 &  0 & -0.473 \\
 $ 0f_{ 7/2}~ 1p_{ 3/2}~ 1p_{ 3/2}~ 0f_{ 5/2}$ &  3 &  0 &  0.142 \\
 $ 0f_{ 7/2}~ 1p_{ 1/2}~ 0f_{ 7/2}~ 1p_{ 1/2}$ &  3 &  0 & -0.628 \\
 $ 0f_{ 7/2}~ 1p_{ 1/2}~ 0f_{ 7/2}~ 0f_{ 5/2}$ &  3 &  0 & -0.279 \\
 $ 0f_{ 7/2}~ 1p_{ 1/2}~ 1p_{ 3/2}~ 1p_{ 3/2}$ &  3 &  0 &  0.544 \\
 $ 0f_{ 7/2}~ 1p_{ 1/2}~ 1p_{ 3/2}~ 0f_{ 5/2}$ &  3 &  0 & -0.119 \\
 $ 0f_{ 7/2}~ 0f_{ 5/2}~ 0f_{ 7/2}~ 0f_{ 5/2}$ &  3 &  0 & -0.525 \\
 $ 0f_{ 7/2}~ 0f_{ 5/2}~ 1p_{ 3/2}~ 1p_{ 3/2}$ &  3 &  0 &  0.439 \\
 $ 0f_{ 7/2}~ 0f_{ 5/2}~ 1p_{ 3/2}~ 0f_{ 5/2}$ &  3 &  0 & -0.369 \\
 $ 1p_{ 3/2}~ 1p_{ 3/2}~ 1p_{ 3/2}~ 1p_{ 3/2}$ &  3 &  0 & -0.826 \\
\end{tabular}
\end{ruledtabular}
\end{table}

\begin{table}[H]
\begin{ruledtabular}
\begin{tabular}{cccc}
\colrule
 $ 1p_{ 3/2}~ 1p_{ 3/2}~ 1p_{ 3/2}~ 0f_{ 5/2}$ &  3 &  0 &  0.227 \\
 $ 1p_{ 3/2}~ 0f_{ 5/2}~ 1p_{ 3/2}~ 0f_{ 5/2}$ &  3 &  0 & -0.258 \\
 $ 0f_{ 7/2}~ 1p_{ 3/2}~ 0f_{ 7/2}~ 1p_{ 3/2}$ &  4 &  0 & -0.106 \\
 $ 0f_{ 7/2}~ 1p_{ 3/2}~ 0f_{ 7/2}~ 1p_{ 1/2}$ &  4 &  0 & -0.302 \\
 $ 0f_{ 7/2}~ 1p_{ 3/2}~ 0f_{ 7/2}~ 0f_{ 5/2}$ &  4 &  0 & -0.229 \\
 $ 0f_{ 7/2}~ 1p_{ 3/2}~ 1p_{ 3/2}~ 0f_{ 5/2}$ &  4 &  0 & -0.620 \\
 $ 0f_{ 7/2}~ 1p_{ 1/2}~ 0f_{ 7/2}~ 1p_{ 1/2}$ &  4 &  0 & -0.593 \\
 $ 0f_{ 7/2}~ 1p_{ 1/2}~ 0f_{ 7/2}~ 0f_{ 5/2}$ &  4 &  0 & -0.590 \\
 $ 0f_{ 7/2}~ 1p_{ 1/2}~ 1p_{ 3/2}~ 0f_{ 5/2}$ &  4 &  0 & -1.006 \\
 $ 0f_{ 7/2}~ 0f_{ 5/2}~ 0f_{ 7/2}~ 0f_{ 5/2}$ &  4 &  0 & -0.843 \\
 $ 0f_{ 7/2}~ 0f_{ 5/2}~ 1p_{ 3/2}~ 0f_{ 5/2}$ &  4 &  0 & -0.782 \\
 $ 1p_{ 3/2}~ 0f_{ 5/2}~ 1p_{ 3/2}~ 0f_{ 5/2}$ &  4 &  0 & -0.756 \\
 $ 0f_{ 7/2}~ 1p_{ 3/2}~ 0f_{ 7/2}~ 1p_{ 3/2}$ &  5 &  0 & -0.916 \\
 $ 0f_{ 7/2}~ 1p_{ 3/2}~ 0f_{ 7/2}~ 0f_{ 5/2}$ &  5 &  0 &  0.282 \\
 $ 0f_{ 7/2}~ 0f_{ 5/2}~ 0f_{ 7/2}~ 0f_{ 5/2}$ &  5 &  0 & -0.184 \\
 $ 0f_{ 7/2}~ 0f_{ 5/2}~ 0f_{ 7/2}~ 0f_{ 5/2}$ &  6 &  0 & -1.570 \\
 $ 0f_{ 7/2}~ 0g_{ 9/2}~ 0f_{ 7/2}~ 0g_{ 9/2}$ &  1 &  0 & -0.843 \\
 $ 0f_{ 7/2}~ 0g_{ 9/2}~ 0f_{ 7/2}~ 1d_{ 5/2}$ &  1 &  0 & -0.191 \\
 $ 0f_{ 7/2}~ 0g_{ 9/2}~ 1p_{ 3/2}~ 1d_{ 5/2}$ &  1 &  0 & -0.674 \\
 $ 0f_{ 7/2}~ 1d_{ 5/2}~ 0f_{ 7/2}~ 1d_{ 5/2}$ &  1 &  0 & -0.636 \\
 $ 0f_{ 7/2}~ 1d_{ 5/2}~ 1p_{ 3/2}~ 1d_{ 5/2}$ &  1 &  0 & -0.248 \\
 $ 1p_{ 3/2}~ 1d_{ 5/2}~ 1p_{ 3/2}~ 1d_{ 5/2}$ &  1 &  0 & -0.847 \\
 $ 0f_{ 7/2}~ 0g_{ 9/2}~ 0f_{ 7/2}~ 0g_{ 9/2}$ &  2 &  0 & -0.590 \\
 $ 0f_{ 7/2}~ 0g_{ 9/2}~ 0f_{ 7/2}~ 1d_{ 5/2}$ &  2 &  0 & -0.454 \\
 $ 0f_{ 7/2}~ 0g_{ 9/2}~ 1p_{ 3/2}~ 1d_{ 5/2}$ &  2 &  0 & -0.405 \\
 $ 0f_{ 7/2}~ 1d_{ 5/2}~ 0f_{ 7/2}~ 1d_{ 5/2}$ &  2 &  0 & -0.443 \\
 $ 0f_{ 7/2}~ 1d_{ 5/2}~ 1p_{ 3/2}~ 1d_{ 5/2}$ &  2 &  0 & -0.437 \\
 $ 1p_{ 3/2}~ 1d_{ 5/2}~ 1p_{ 3/2}~ 1d_{ 5/2}$ &  2 &  0 & -0.488 \\
 $ 0f_{ 7/2}~ 0g_{ 9/2}~ 0f_{ 7/2}~ 0g_{ 9/2}$ &  3 &  0 & -0.179 \\
 $ 0f_{ 7/2}~ 0g_{ 9/2}~ 0f_{ 7/2}~ 1d_{ 5/2}$ &  3 &  0 & -0.229 \\
 $ 0f_{ 7/2}~ 0g_{ 9/2}~ 1p_{ 3/2}~ 0g_{ 9/2}$ &  3 &  0 & -0.436 \\
 $ 0f_{ 7/2}~ 0g_{ 9/2}~ 1p_{ 3/2}~ 1d_{ 5/2}$ &  3 &  0 & -0.218 \\
 $ 0f_{ 7/2}~ 1d_{ 5/2}~ 0f_{ 7/2}~ 1d_{ 5/2}$ &  3 &  0 & -0.108 \\
 $ 0f_{ 7/2}~ 1d_{ 5/2}~ 1p_{ 3/2}~ 0g_{ 9/2}$ &  3 &  0 & -0.211 \\
 $ 0f_{ 7/2}~ 1d_{ 5/2}~ 1p_{ 3/2}~ 1d_{ 5/2}$ &  3 &  0 & -0.213 \\
 $ 1p_{ 3/2}~ 0g_{ 9/2}~ 1p_{ 3/2}~ 0g_{ 9/2}$ &  3 &  0 & -0.772 \\
 $ 1p_{ 3/2}~ 0g_{ 9/2}~ 1p_{ 3/2}~ 1d_{ 5/2}$ &  3 &  0 & -0.378 \\
 $ 1p_{ 3/2}~ 1d_{ 5/2}~ 1p_{ 3/2}~ 1d_{ 5/2}$ &  3 &  0 & -0.250 \\
 $ 0f_{ 7/2}~ 0g_{ 9/2}~ 0f_{ 7/2}~ 0g_{ 9/2}$ &  4 &  0 & -0.070 \\
 $ 0f_{ 7/2}~ 0g_{ 9/2}~ 0f_{ 7/2}~ 1d_{ 5/2}$ &  4 &  0 & -0.351 \\
 $ 0f_{ 7/2}~ 0g_{ 9/2}~ 1p_{ 3/2}~ 0g_{ 9/2}$ &  4 &  0 & -0.397 \\
 $ 0f_{ 7/2}~ 0g_{ 9/2}~ 1p_{ 3/2}~ 1d_{ 5/2}$ &  4 &  0 & -0.357 \\
 $ 0f_{ 7/2}~ 1d_{ 5/2}~ 0f_{ 7/2}~ 1d_{ 5/2}$ &  4 &  0 & -0.254 \\
 $ 0f_{ 7/2}~ 1d_{ 5/2}~ 1p_{ 3/2}~ 0g_{ 9/2}$ &  4 &  0 & -0.329 \\
 $ 0f_{ 7/2}~ 1d_{ 5/2}~ 1p_{ 3/2}~ 1d_{ 5/2}$ &  4 &  0 & -0.483 \\
 $ 1p_{ 3/2}~ 0g_{ 9/2}~ 1p_{ 3/2}~ 0g_{ 9/2}$ &  4 &  0 & -0.190 \\
 $ 1p_{ 3/2}~ 0g_{ 9/2}~ 1p_{ 3/2}~ 1d_{ 5/2}$ &  4 &  0 & -0.409 \\
 $ 1p_{ 3/2}~ 1d_{ 5/2}~ 1p_{ 3/2}~ 1d_{ 5/2}$ &  4 &  0 & -0.770 \\
 $ 0f_{ 7/2}~ 0g_{ 9/2}~ 0f_{ 7/2}~ 0g_{ 9/2}$ &  5 &  0 & -0.015 \\
 $ 0f_{ 7/2}~ 0g_{ 9/2}~ 0f_{ 7/2}~ 1d_{ 5/2}$ &  5 &  0 & -0.147 \\
 $ 0f_{ 7/2}~ 0g_{ 9/2}~ 1p_{ 3/2}~ 0g_{ 9/2}$ &  5 &  0 & -0.297 \\
 $ 0f_{ 7/2}~ 1d_{ 5/2}~ 0f_{ 7/2}~ 1d_{ 5/2}$ &  5 &  0 & -0.029 \\
 $ 0f_{ 7/2}~ 1d_{ 5/2}~ 1p_{ 3/2}~ 0g_{ 9/2}$ &  5 &  0 & -0.134 \\
 $ 1p_{ 3/2}~ 0g_{ 9/2}~ 1p_{ 3/2}~ 0g_{ 9/2}$ &  5 &  0 & -0.069 \\
 $ 0f_{ 7/2}~ 0g_{ 9/2}~ 0f_{ 7/2}~ 0g_{ 9/2}$ &  6 &  0 & -0.125 \\
 $ 0f_{ 7/2}~ 0g_{ 9/2}~ 0f_{ 7/2}~ 1d_{ 5/2}$ &  6 &  0 & -0.447 \\
 $ 0f_{ 7/2}~ 0g_{ 9/2}~ 1p_{ 3/2}~ 0g_{ 9/2}$ &  6 &  0 & -0.538 \\
 $ 0f_{ 7/2}~ 1d_{ 5/2}~ 0f_{ 7/2}~ 1d_{ 5/2}$ &  6 &  0 & -0.742 \\
 $ 0f_{ 7/2}~ 1d_{ 5/2}~ 1p_{ 3/2}~ 0g_{ 9/2}$ &  6 &  0 & -0.671 \\
 $ 1p_{ 3/2}~ 0g_{ 9/2}~ 1p_{ 3/2}~ 0g_{ 9/2}$ &  6 &  0 & -0.479 \\
 $ 0f_{ 7/2}~ 0g_{ 9/2}~ 0f_{ 7/2}~ 0g_{ 9/2}$ &  7 &  0 &  0.063 \\
 $ 0f_{ 7/2}~ 0g_{ 9/2}~ 0f_{ 7/2}~ 0g_{ 9/2}$ &  8 &  0 & -1.119 \\
\end{tabular}
\end{ruledtabular}
\end{table}

\begin{table}[H]
\caption{Proton effective charges of the electric quadrupole operator
  $E2$ for the model spaces (I) and (II).}
\begin{ruledtabular}
\begin{tabular}{ccc}
$n_a l_a j_a ~ n_b l_b j_b $ &  $\langle a || e_n || b \rangle $ (I) &  $\langle a || e_n || b \rangle $ (II) \\
\colrule
 $0f_{ 7/2}~ 0f_{ 7/2}$  &   1.22  &   1.23 \\
 $0f_{ 7/2}~ 1p_{ 3/2}$  &   1.34  &   1.36 \\
 $1p_{ 3/2}~ 1p_{ 3/2}$  &   1.40  &   1.42 \\
\end{tabular}
\end{ruledtabular}
\label{tableep}
\end{table}

\begin{table}[H]
\caption{Neutron effective charges of the
  electric quadrupole operator $E2$ for the model spaces (I) and (II).}
\begin{ruledtabular}
\begin{tabular}{ccc}
$n_a l_a j_a ~ n_b l_b j_b $ &  $\langle a || e_n || b \rangle $ (I) &  $\langle a || e_n || b \rangle $ (II) \\
\colrule
 $1p_{ 3/2}~ 1p_{ 3/2}$  &   0.38  &   0.40 \\
 $1p_{ 3/2}~ 1p_{ 1/2}$  &   0.39  &   0.42 \\
 $1p_{ 3/2}~ 0f_{ 5/2}$  &   0.43  &   0.46 \\
 $1p_{ 1/2}~ 0f_{ 5/2}$  &   0.41  &   0.45 \\
 $0f_{ 5/2}~ 0f_{ 5/2}$  &   0.59  &   0.61 \\
 $0g_{ 9/2}~ 0g_{ 9/2}$  &   0.39  &   0.40 \\
 $0g_{ 9/2}~ 1d_{ 5/2}$  &    ~    &   0.35 \\
 $1d_{ 5/2}~ 1d_{ 5/2}$  &    ~    &   0.36 \\
\end{tabular}
\end{ruledtabular}
\label{tableen}
\end{table}

\end{document}